\documentclass[11pt]{article}

\usepackage{graphicx}
\usepackage[labelsep=quad,indention=10pt]{subfig}
\usepackage{wrapfig}
\usepackage[english]{babel}
\usepackage{xspace}
\usepackage{amsthm}
\usepackage{amsmath}
\usepackage{amsfonts}
\newtheorem{theorem}{Theorem}

\usepackage{tikz}
\usetikzlibrary{fadings}
\usetikzlibrary{patterns}
\usetikzlibrary{shadows.blur}
\usetikzlibrary{shapes}
\input{extensionAlgis.sty}
\usepackage{pgfplots} 
\pgfplotsset{compat=1.18}
\usepackage{pgfplotstable}
\usetikzlibrary{intersections}
\usepgfplotslibrary{fillbetween}
\usetikzlibrary{patterns}
\usepackage{orcidlink}

\usepackage{hyperref}
\hypersetup{
    colorlinks=true,
    linkcolor=bleuF,
    citecolor=bleuF,
    filecolor=magenta,      
    urlcolor=bleuF,
    pdftitle={Arc Blanc: a real time ocean simulation framework},
    pdfpagemode=FullScreen,
}
\usepackage{authblk}
\usepackage{algorithm}
\usepackage{algpseudocode}
\algrenewcommand{\algorithmiccomment}[1]{\hfill \textcolor{greenF}{// #1}}

\usepackage{multirow}

\def\Framework{The Arc Blanc framework\xspace}
\def\framework{the Arc Blanc framework\xspace}
\def\etal{\emph{et al.}\xspace}
\def\ie{\emph{i.e.}\xspace}

\title{Arc Blanc: a real time ocean simulation framework}
\author[1,2,*]{David Algis \orcidlink{0009-0004-6033-2389}}
\author[3]{Berenger Bramas \orcidlink{0000-0003-0281-9709}}
\author[1]{Emmanuelle Darles \orcidlink{0000-0002-6172-1237}}
\author[1]{Lilian Aveneau \orcidlink{0000-0003-3129-1149}}

\affil[1]{Université de Poitiers, Univ. Limoges, CNRS, XLIM, France}
\affil[2]{Studio Nyx}
\affil[3]{Inria Nancy Grand Est, ICube, France}

\affil[*]{%
  Corresponding Author : \href{mailto:david.algis@univ-poitiers.fr}{david.algis@univ-poitiers.fr}
}
\date{2025-02-13}

\begin{document}

\maketitle

\begin{figure}[h]
	\centering
    \includegraphics[width=4in]{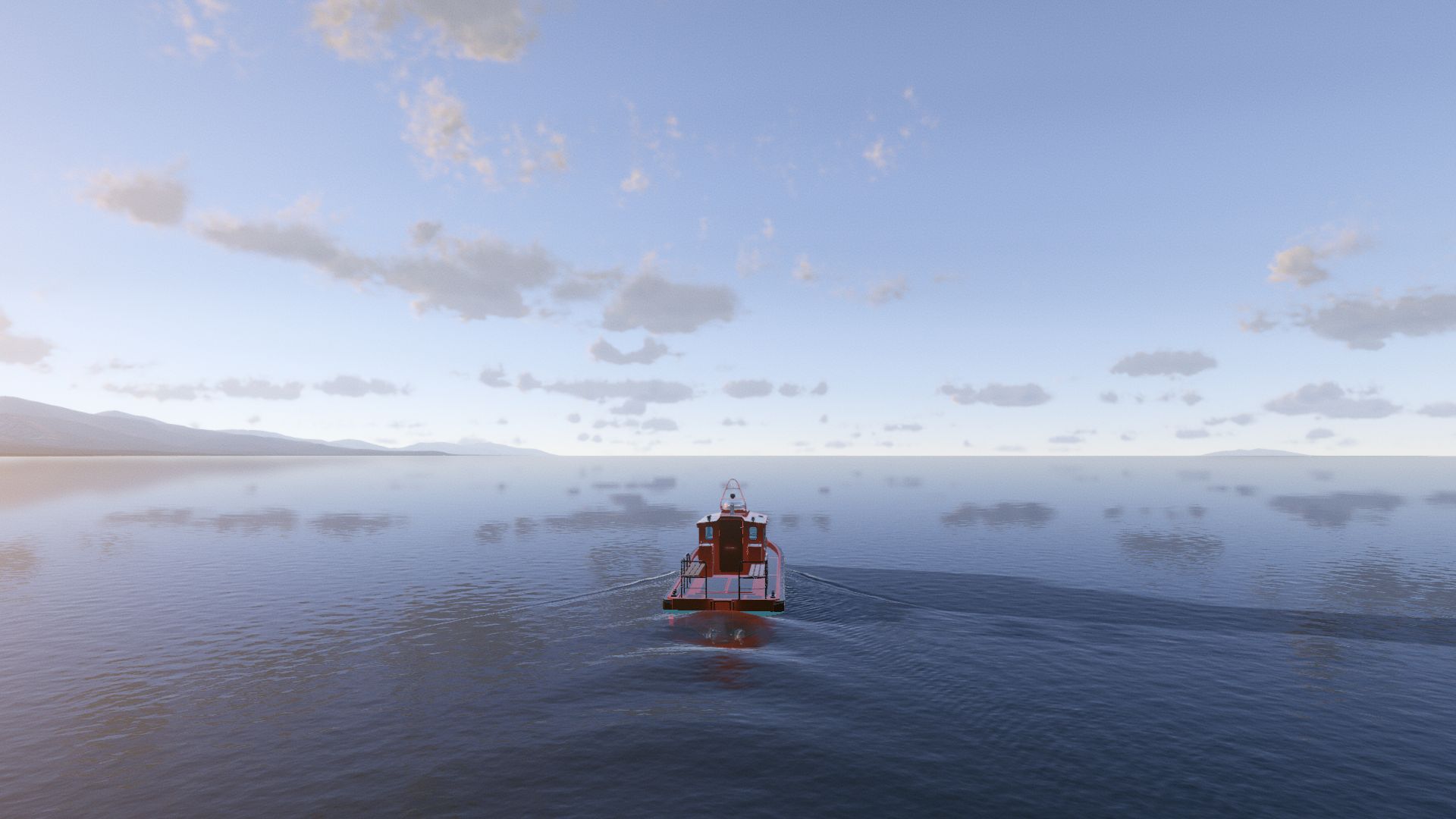}
    \caption{Example of rendering of \framework.}
    \label{fig:teaser}
\end{figure}

\begin{abstract}
	\small
	The oceans cover the vast majority of the Earth.
	Therefore, their simulation has many scientific, industrial and military interests, including computer graphics domain.
	By fully exploiting the multi-threading power of GPU and CPU, current state-of-the-art tools can achieve real-time ocean simulation, even if it is sometimes needed to reduce the physical realism for large scenes.
	Although most of the building blocks for implementing an ocean simulator are described in the literature, a clear explanation of how they interconnect is lacking.
	Hence, this paper proposes to bring all these components together, detailing all their interactions, in a comprehensive and fully described real-time framework that simulates the free ocean surface and the coupling between solids and fluid.
	This article also presents several improvements to enhance the physical realism of our model.
	The two main ones are:
	calculating the real-time velocity of ocean fluids at any depth;
	computing the input of the solid to fluid coupling algorithm.
\end{abstract}

\section{Introduction}
\label{sec:introduction}

\subsection{Context}
\label{subsec:intro-context}
Simulating the dynamics of an ocean and its surrounding elements remains an important objective in computer graphics.
As all the oceans encompass a substantial portion of the Earth's surface, understanding their behavior is of paramount importance in scientific, industrial, entertainment and military fields.

The computational power of modern GPUs and CPUs, coupled with the latest high-speed memory, has greatly simplified the processing of fluid dynamics simulations.
They make it possible to perform real-time simulations on large discrete grids surrounded by solids, albeit with some physical approximations, thanks to their massively parallel capabilities.

\begin{figure}
	\centering
	\tikzset{every picture/.style={line width=0.75pt}} 
	\input{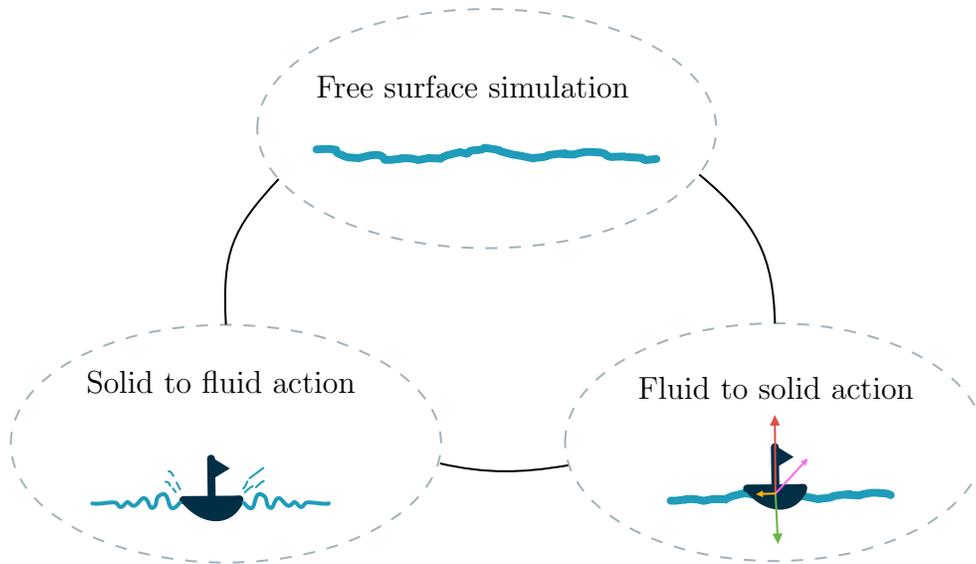}
	\caption{Overview of \framework components: free surface, solid to fluid and fluid to solid coupling.}
	\label{fig:overviewFramework}
\end{figure}

While there are many articles describing different parts of ocean simulation, the literature lacks papers that bring all the pieces together.
This article proposes a fully described real-time framework designed to simulate the free surface of the ocean and the interactions between solid objects and the surrounding fluid in real time (see Figure \ref{fig:overviewFramework}).
This new framework incorporates three different methods from the literature:
\begin{itemize}
	\item The simulation of the free surface of the ocean, initially developed by Tessendorf (2001) \cite{tessendorfSimulatingOceanWater2001} and its improvements along the years \cite{horvathEmpiricalDirectionalWave2015,tessendorfGilliganPrototypeFramework2017}.
	\item The estimation of the forces applied by the fluid on a solid mesh, mixing together different techniques \cite{yukselRealtimeWaterWaves2010,kellomakiRigidBodyInteraction2014}.
	\item The simulation of the alteration of a free surface by a solid, inspired by Cords and Staadt works (2009) \cite{cordsRealTimeOpenWater2009}.
\end{itemize}
To bring together these previous works, this new framework proposes two new contributions:
\begin{itemize}
	\item An algorithm to compute the ocean velocity at any depth from the simulation of the free surface. This velocity is applied to the fluid to solid forces.
	\item An algorithm to compute the input of the Cords and Staadt method more realistically from the free surface.
\end{itemize}

This article is organized as follows.
Section~\ref{sec:relatedwork} presents the previous works related to ocean simulation, including its interaction with solid.
Sections \ref{sec:oceanWaves} to \ref{sec:solidToFluidAction} detail \framework, including the ocean surface simulation in Section~\ref{sec:oceanWaves}, the action of the fluid to solids in Section~\ref{sec:fluidToSolidAction} and the action of a solid to the fluid in Section~\ref{sec:solidToFluidAction}.
Section~\ref{sec:results_and_discussions} discusses some results obtained with \framework.
At last, Section~\ref{sec:conclusions} gives general conclusions.

\section{Related works}
\label{sec:relatedwork}
This section presents the main works related to the simulation of ocean, including the free surface, the fluid to solid and solid to fluid coupling.

\subsection{Ocean free surface simulation}
\label{subsubsec:oceanFreeSurfaceSim}
J. Tessendorf (2001) \cite{tessendorfSimulatingOceanWater2001} presented what could be considered a seminal work in real-time ocean simulation. His work relies on the Fast Fourier Transform (FFT) and statistical wave spectra to accurately find the analytical solution to an approximation of the Navier-Stokes equations on a 2D height field. This method is commonly known as the “Tessendorf method”.

The realism of this method was significantly improved by Horvath (2015) \cite{horvathEmpiricalDirectionalWave2015}, who proposed an empirically based directional spectrum.
In addition, he proposed to add a parameter called “swell”, which allows the transition from empirical directional propagation to parallel waves.

Another notable improvement of Tessendorf's method is the reduction of artifact effects through the use of multiple displacement layers.
This technique, introduced by Dupuy and Bruneton (2012) \cite{dupuyRealtimeAnimationRendering2012}, involves stacking multiple displacement layers, each one representing different scales of wave motion.
By combining these layers, the artifacts and repetitive patterns that can often occur in simulated water surfaces are effectively reduced or eliminated.
Finally, Tessendorf (2017) \cite{tessendorfGilliganPrototypeFramework2017} summarized these works in a prototype framework for large-scale water simulation.

\Framework aggregates all these methods in order to simulate the ocean free surface.
\subsection{Fluid to solid coupling}
\label{subsubsec:fluidToSolidCoupling}
There are several ways to describe the motion of a solid in a given fluid.
Many particle-based methods have been developed in recent years, as presented by Koshier \etal (2022) \cite{koschierSurveySPHMethods2022}.
Akinci \etal (2012) \cite{akinciVersatileRigidfluidCoupling2012} have proposed to calculate contact forces between fluid particles and boundary ones in combination with Predictive-Corrective Incompressible Smoothed Particles Hydrodynamics (PCISPH) \cite{solenthalerPredictivecorrectiveIncompressibleSPH2009}.
Ihmsen \etal (2014) \cite{ihmsenImplicitIncompressibleSPH2014} have applied the same principle to Implicit Incompressible SPH (IISPH), and Bender and Koschier (2017) \cite{benderDivergenceFreeSPHIncompressible2017} to Divergence Free SPH (DFSPH).
These works have been improved and stabilized by Gissler  \etal \cite{gisslerInterlinkedSPHPressure2019}.
These methods look promising because of their physical accuracy.
Nevertheless, their performance is linked to the number of particles used for the simulation, which makes them impossible to apply in real time for large scale environments.

Another approach related to the dynamics of a ship in water is the calculation of the so-called \emph{Response Amplitude Operator} (RAO), which is expressed using strip theory by Salvesen \etal (1970) \cite{salvesenShipMotionsSea1970}. Fonseca and Guedes Soares (1998) \cite{fonsecaTimeDomainAnalysisLargeAmplitude1998} improved the method to support more nonlinear effects and large amplitude motions. Varela and Guedes Soares (2011) \cite{varelaInteractiveSimulationShip2011} have used a table of pre-calculated RAO, which are later interpolated to achieve an interactive frame rate.
However, to our knowledge, this type of method is either too expensive for real-time simulation or not general enough because it requires some physical pre-computation based on computational fluid dynamics or experimental data.

Finally, another attempt to achieve real-time performance is to approximate the forces induced by the fluid on the solid on a mesh.
Basically this is a less accurate method, but it's more efficient and general as it avoids significant pre-calculation.
Yuksel (2010) \cite{yukselRealtimeWaterWaves2010} in his thesis proposed a set of three forces: buoyancy, drag and lift.
Kellomaki (2014) \cite{kellomakiRigidBodyInteraction2014} focused only on buoyancy and drag forces.
Casas-Yrurzum \etal \cite{casas-yrurzumRealtimePhysicsSimulation2012a} simulates the motion of speedboats by decomposing the solid as a set of small cubes.
Kerner (2016) \cite{kernerWaterInteractionModel2016} wrote a full article on boat physics for video games, adding more forces to simulate slamming and linear damping.
\Framework relies on these forces approximation, aggregating these four last papers.

\subsection{Solid to fluid coupling}
\label{subsubsec:solidToFluidCoupling}
The simulation of the surface of water in contact with a solid body has been an active research topic for the last twenty years.
Gomez (2000) \cite{gomezInteractiveSimulationWater2000} solved the wave equation using the finite-difference method (FDM) to simulate waves on the free surface.
This method was improved by Cords and Staadt (2009) \cite{cordsRealTimeOpenWater2009} to work with moving objects.
Tessendorf (2004) \cite{tessendorfInteractiveWaterSurface2004} proposed a convolutional method.
He extended his method by using an exponential solver \cite{tessendorfEWaveUsingExponential2014}.
Yuksel (2007) \cite{yukselWaveParticles2007} used a Lagrangian paradigm to model the waves as particles, which he called wave-particles.
Chentanez and Müller (2010) \cite{chentanezRealTimeSimulationLarge2010} created a hybrid solver combining Eulerian and Lagrangian approaches.
Canabal \etal (2016) \cite{canabalDispersionKernelsWater2016} improved the physical accuracy of the convolutional method by introducing a dispersion kernel.
Jeschke and Woktan (2017) \cite{jeschkeWaterWavePackets2017} took the work of C. Yuskel a bit further by introducing the notion of a wave packet, which carries information about an entire wave train.
One year later, Jeschke \etal (2018) \cite{jeschkeWaterSurfaceWavelets2018} proposed to use a wavelet transformation that discretizes the wave amplitudes as a function of space, frequency and direction.
Finally, Schreck \etal (2019) \cite{schreckFundamentalSolutionsWater2019} based their work on the method of fundamental solutions to generate ambient waves on a large scale domain interacting with complex boundaries.

For solid-fluid coupling, \framework based its calculations on the FDM described in Gomez (2000) \cite{gomezInteractiveSimulationWater2000} and its improvement by Cords and Staadt (2009) \cite{cordsRealTimeOpenWater2009}, because its ability to be real-time while providing quite realistic behavior.

\section{Ocean waves}
\label{sec:oceanWaves}
This section summarizes Tessendorf's method and the various improvements it has received.
It also presents our first contribution: calculation of the fluid velocity at any depth.

Let us introduce some notations as follows:
\begin{itemize}
	\item The scalar $H\in\mathbb R^+$ denotes the depth of the water. \Framework assumes \textbf{deep water} and therefore supposes that $H$ tends towards $+\infty$\footnote{See Appendix \ref{sec:annexe} for more details on this assumption.}.
	\item The scalar $h(\mathbf x, t)\in\mathbb R$ is the height of the water surface at the horizontal position $\mathbf x=\left[x~z\right]^T$ and at time $t$.
	\item The vector $\mathbf{v}(\mathbf x, y, t) =  \left[v_x(\mathbf x, y, t)\ v_y(\mathbf x, y, t)\ v_z(\mathbf x, y, t)\right]^T\in\mathbb R^3$ represents the velocity of the fluid at the horizontal position $\mathbf x$, depth $y$ and time $t$.
	\item The scalar $\phi(\mathbf x,y,t)\in\mathbb R$ is the potential velocity of fluid at the horizontal position $\mathbf x$, depth $y$ and time $t$.
	\item The scalar $g=9.80665~m.s^{-2}$ denotes the gravitational acceleration, while $\mathbf g\in\mathbb R^3$ is the gravitational field.
\end{itemize}

Tessendorf method is based on the following approximations of the Navier-Stokes equations\footnote{See Appendix  \ref{sec:annexe-NS}.}:
\begin{equation}
	\label{equa:systemTessendorf}
	\left\lbrace
	\begin{array}{ll}
		\frac{\partial\phi}{\partial t} = -gh(\mathbf x,t)            & \text{for }y=0,            \\
		\Delta\phi=0                                                  & \text{for }-H\leq y\leq 0, \\
		\frac{\partial h}{\partial t}=\frac{\partial\phi}{\partial y} & \text{for }y=0,            \\
		\frac{\partial\phi}{\partial y}=0                             & \text{for }y=-H.
	\end{array}
	\right.
\end{equation}

\subsection{Water height}
\label{subsec:waterHeight}
Tessendorf (2001) \cite{tessendorfSimulatingOceanWater2001} proposed to solve Equation~\ref{equa:systemTessendorf} using a sum of waves of different amplitudes and wavelengths as follows:
\begin{equation}
	\label{equa:heightWater}
	h(\mathbf x,t)=\sum_{\mathbf{k}} \tilde{h}(\mathbf{k}, t) \exp\left( i \mathbf{k}\cdot \mathbf x\right),
\end{equation}
where $\mathbf{k}=\left[k_x\ k_z\right]^T$ is the wave vector with $k_x=\frac{2\pi n}{L_{i}}$ and $k_z=\frac{2\pi m}{L_{i}}$, $L_i$ is the length of the $i$-th cascade as defined in Section~\ref{subsec:cascades}, $n$ and $m$ are integers in $\left[-\frac{N}{2},\frac{N}{2}\right]$, $N\in\mathbb N^+$ being a constant given by user, and $\tilde{h}$ is given by:
\begin{equation}
	\label{equa:heightWaterBis}
	\tilde{h}(\mathbf{k}, t) = \tilde{h}_0(\mathbf{k})\exp(i\omega(k)t)+\tilde{h}_0^*(-\mathbf{k})\exp(-i\omega(k)t),
\end{equation}
where $k$ is the Euclidean norm of the wave vector $\mathbf{k}$, $\omega(k)=\sqrt{gk}$ is the \textbf{dispersion relation}\footnote{The dispersion relation is defined for deep water, see Appendix~\ref{sec:annexe} for more details.} and $\tilde{h}_0^*$ the conjugate of $\tilde{h}_0$.
Researchers in oceanography have made statistical measurements on the ocean to compute $\tilde{h}_0$, leading to many different \textit{ocean spectra} \cite{tessendorfGilliganPrototypeFramework2017}.
In its original paper, Tessendorf (2001) \cite{tessendorfSimulatingOceanWater2001} used the \textit{Phillips Spectrum} with Gaussian fluctuation, leading to the following expression of $\tilde{h}_0(\mathbf{k})$:
\begin{equation}
	\tilde{h}_0(\mathbf{k}) = \xi_{\mathbf{k}}\sqrt{\frac{4\pi}{L_ik}S(\omega)D(\omega, \theta)\left|\frac{\partial\omega(k)}{\partial k}\right|}
\end{equation}
where $\xi_{\mathbf{k}}$ are samples from a Gaussian distribution with mean $0$ and standard deviation $1$, $S(\omega)$ and $D(\omega, \theta)$ are the frequency and directional spectra described respectively in paragraphs \ref{subsubsec:jonswap} and \ref{subsubsec:donelanbanner}.

\subsubsection{JONSWAP frequency spectrum}
\label{subsubsec:jonswap}

\Framework uses the frequency spectrum developed by Hasselmann \etal (1973) \cite{k.hasselmannMeasurementsWindwaveGrowth1973} in the Joint North Sea Wave Project (JONSWAP).
This spectrum was developed for deep sea water, and hence is in line with our deep water hypothesis.
It is expressed as follows:
\begin{equation}
	S(\omega)=\frac{\alpha g^2}{\omega^5} \exp\left(-\frac{5}{4}\left(\frac{\omega_p}{\omega}\right)^4\right)\gamma^r
\end{equation}
where:
\begin{itemize}
	\item The constant $\alpha=0.076\left(\frac{U^2}{F g}\right)^{0.22}$, $U$ being the wind speed in $m.s^{-1}$ and $F$ the fetch\footnote{The fetch is the length of the area over which the wind is acting on the water, that is to say the distance from a lee shore.} in $m$.
	\item The constant $\omega_p=22\left(\frac{g^2}{U F}\right)$.
	\item The constant $\gamma=3.3$.
	\item The exponent term $r$ is defined as follows:
	      \begin{equation}
		      r=\exp\left(-\frac{\left(\omega-\omega_p\right)^2}{2 \sigma^2 \omega_p^2}\right)
	      \end{equation}
	\item The term $\sigma$ is as follows:
	      \[
		      \sigma=\begin{cases}
			      0.07 & \omega \leq \omega_p, \\
			      0.09 & \omega>\omega_p.
		      \end{cases}
	      \]
\end{itemize}

\subsubsection{Custom Donelan-Banner spectrum}
\label{subsubsec:donelanbanner}

In \framework, the directional spectrum relies on the work of Horvath (2015) \cite{horvathEmpiricalDirectionalWave2015}, that mixes together the Donelan-Banner spectrum to control the direction of the waves and a “swell parameter”.
Let us recall that, if the wind direction is given by $\theta_0$, the wind direction bias $\theta$ can be defined as follows:
\begin{equation}
	\theta = atan2(k_z, k_x) - \theta_0
\end{equation}
Moreover, the Donelan-Banner $D_{DB}$ spectrum is given as follows:
\begin{equation}
	D_{DB}\left(\omega,\theta\right)=\frac{1}{2}Q_{DB}\left(\omega\right)\beta_s\left(\operatorname{sech}\left(\beta_s\theta\right)\right)^2.
	\label{equa:ddb}
\end{equation}

This expression depends on the ratio $r_{\omega}=\omega/\omega_p$ of $\omega$ and $\omega_p$, the latter being the same as in JONSWAP spectrum.
Hence, in \framework $\beta_s$ is defined as a mix of Donelan \etal (1985) \cite{donelanDirectionalSpectraWindgenerated1985} for $r_{\omega}<1.6$, and Banner (1990) \cite{bannerEquilibriumSpectraWind1990} for $r_{\omega}\geq1.6$.
Moreover, the case $r_{\omega}<0.56$ is removed, as suggested by C. Horvath.
Altogether, this leads to the following expression of $\beta_s$:
\begin{equation}
	\begin{aligned}
		 & \beta_s= \begin{cases}%
			            2.61\left(r_{\omega}\right)^{1.3}  & \text{for } r_{\omega}<0.95          \\
			            2.28\left(r_{\omega}\right)^{-1.3} & \text{for } 0.95 \leq r_{\omega}<1.6 \\
			            10^\epsilon                        & \text{for }  r_{\omega}\geq1.6       %
		            \end{cases}
	\end{aligned}
\end{equation}
where
\begin{equation}
	\epsilon=0.8393 \exp \left(-0.567\ln\left(r_{\omega}^2\right)\right)-0.4.
\end{equation}
Any directional spectrum $D(\omega,\theta)$ must satisfy the normalization condition:
\begin{equation}
	\label{equa:normalization}
	\int_{-\pi}^\pi D(\omega,\theta)d\theta = 1
\end{equation}
Hence, $Q_{DB}(\omega)$ is a normalization factor to make $D_{DB}(\omega, \theta)$ respect condition \ref{equa:normalization}:
\begin{equation}
	\begin{aligned}
		Q_{DB}(\omega) & = \frac{1}{\displaystyle\int_{-\pi}^\pi \frac{1}{2}\beta_s\left(\operatorname{sech}\left(\beta_s \theta\right)\right)^2 d\theta} = \frac{1}{\tanh\left(\beta_s\pi\right)}
	\end{aligned}
\end{equation}

\Framework simulates swell, corresponding to more elongated waves that have traveled out of their generating area and that mixes with wind-generated local waves.
As in Horvath (2015) \cite{horvathEmpiricalDirectionalWave2015}, it relies on the swell parameter $\xi$ related to the spectrum $D_{\xi}$, defined as follows:
\begin{equation}
	D_{\xi}(\omega, \theta)=|\cos (\theta / 2)|^{2 s_{\xi}}
\end{equation}
where the spreading swell function $s_\xi$ is defined by:
\begin{equation}
	s_\xi = 16 \tanh(1 / r_\omega) \xi^2
\end{equation}
Note that, with $\xi=0$, the spectrum $D_{0}(\omega, \theta)=\frac{1}{2\pi}$ is constant while for $\xi=1$ the waves become much more elongated.

The product between the Donelan-Banner and $D_{\xi}$ spectra is denoted by $D_{DB\xi}$, and is expressed as follows:
\begin{equation}
	D_{DB\xi}(\omega, \theta) =Q_{DB\xi}(\omega) D_{DB}(\omega, \theta) D_{\xi}(\omega, \theta)
\end{equation}
where the normalization factor is:
\begin{equation}
	Q_{DB\xi}(\omega)=\frac{1}{\displaystyle\int_{-\pi}^\pi D_{DB}(\omega, \theta) D_{\xi}(\omega, \theta) \mathrm{d} \theta}.
\end{equation}
This normalization factor $Q_{DB\xi}(\omega)$ is computationally expensive.

It can be noticed that it depends only on the ratio $r_{\omega}$.
Hence, \framework uses a simple method to approximate it, through interpolations relying on multiple Lagrangian polynomials as follows:
\begin{equation}
	\widetilde{Q_{DB\xi}}(r_{\omega})=
	\begin{cases}
		7.1467551 r_{\omega}^2 - 13.4662001 r_{\omega} + 7.75651088    & \text{for }r_{\omega} < 0.94,        \\
		- 0.69906109 r_{\omega}^2 + 0.77975933 r_{\omega} + 0.10169164 & \text{for }0.94 \leq r_{\omega} < 5, \\
		- 2.1860997 r_{\omega}^2 +0.0269209 r_{\omega} +0.00016283     & \text{for } 5 \leq r_{\omega} < 100, \\
		1.2038847 r_{\omega} + 0.0008147                               & \text{else.}
	\end{cases}
\end{equation}

Computed using $10000$ random samples, the normalization integral (Equation \ref{equa:normalization}) of such an approximation is $1.0699$ in average.

Note that a first normalization factor $Q_{DB}$ is applied to the Donelan-Banner spectrum (Equation \ref{equa:ddb}) before calculating the second normalization factor $\widetilde{Q_{DB\xi}}(r_{\omega})$. In fact, a normalization problem will occur if the full-spectrum normalization factor is simply approximated before its application, because for small values of $r_{\omega}$, $Q_{DB\xi}(\omega)$ will be very low and therefore numerically challenging to normalize.

\begin{figure}[htbp]
	\centering

	\subfloat[Using $\delta = 0$.]
	{
		\includegraphics[width=0.45\textwidth]{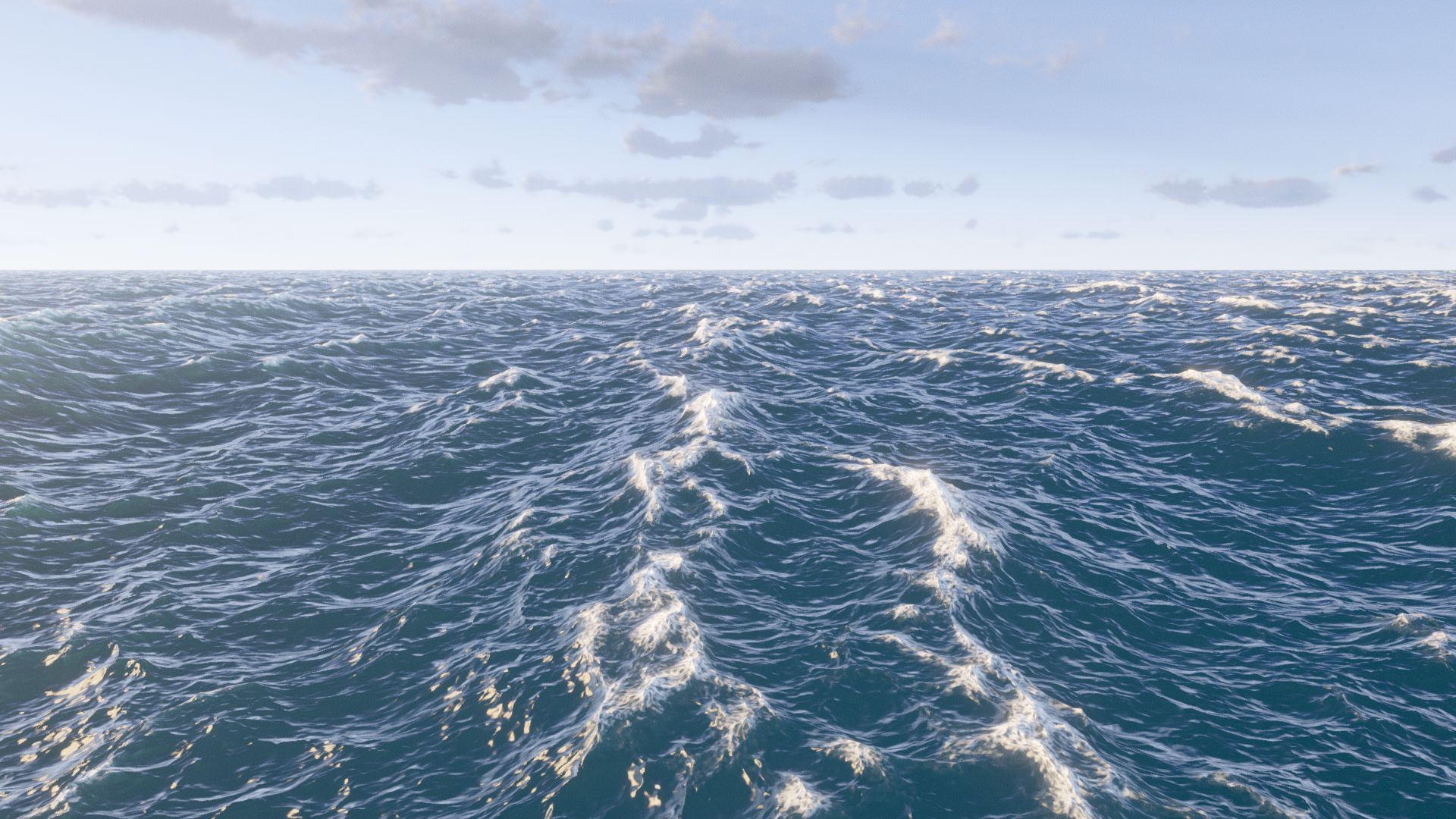}
		\label{fig:dispersion0}
	}
	\hfill
	\subfloat[Using $\delta = 1$.]
	{
		\includegraphics[width=0.45\textwidth]{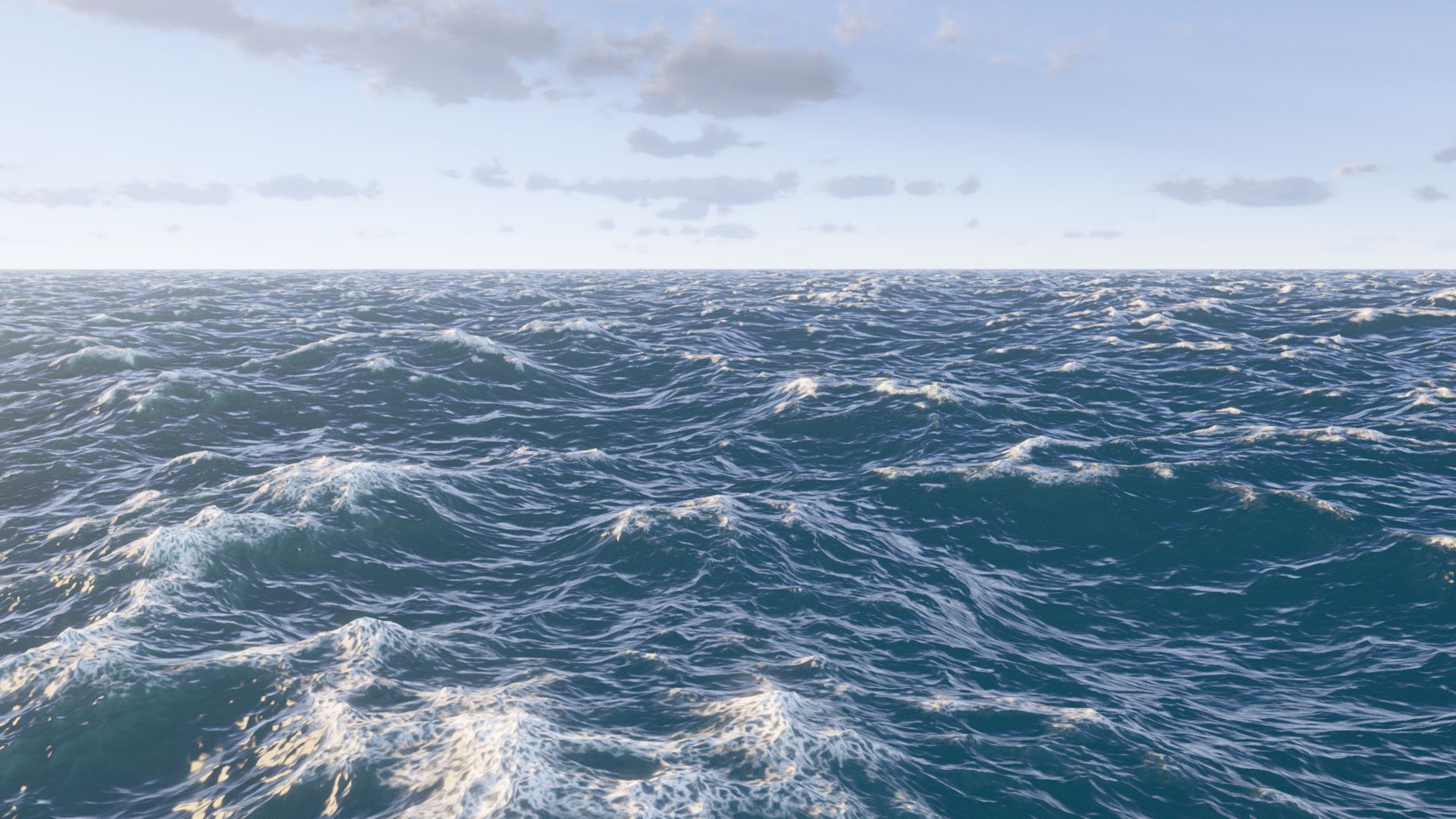}
		\label{fig:dispersion1}
	}
	\caption{Effect of the parameter $\delta$ on sea state $5$ on the Beaufort scale.}
	\label{fig:dispersionEffect}
\end{figure}

Finally, \framework uses a last parameter $\delta\in[0,1]$ that represents how much the waves orientation is constant ($\delta = 0$) or not ($\delta = 1$) (see Figure~\ref{fig:dispersionEffect}).
Its meaning is as follows:
\begin{itemize}
	\item $\delta = 0$ indicates a \emph{neutral} directional spectrum, that has no consequence on $\tilde{h}_0$;
	\item $\delta=1$ indicates a spectrum equals to $D_{DB\xi}(\omega, \theta)$.
\end{itemize}
This gives the final expression of the custom Donelan-Banner spectrum used in \framework:
\begin{equation}
	D(\omega, \theta)=\left(1-\delta\right)\frac{1}{2\pi}+\delta D_{DB\xi}(\omega,\theta).
\end{equation}

\subsection{Water displacement and derivative}
\label{subsec:water_displacement_and_derivative}
\Framework uses the horizontal displacement $\mathbf{D}(\mathbf x,t) = \left[D_x(\mathbf x,t)\ D_z(\mathbf x,t)\right]^T$ of seawater to simulate more realistic choppy waves.
As proposed in Sections 4.5 and 4.6 of Tessendorf (2001) \cite{tessendorfSimulatingOceanWater2001}, it is calculated as follows:
\begin{equation}\label{equa:hori-displacement}
	\mathbf{D}(\mathbf x,t) = \sum_{\mathbf{k}} \mathbf{\tilde{D}}(\mathbf k,t)\exp\left(i\mathbf{k}\cdot\mathbf x\right)
\end{equation}

where
\begin{equation}\label{equa:hori-displacement-k}
	\mathbf{\tilde{D}}(\mathbf k,t) = \left[\tilde{D}_x(\mathbf k,t)\ \tilde{D}_z(\mathbf k,t)\right]^T = \frac{i\mathbf{k}}{k} \tilde{h}(\mathbf{k}, t).
\end{equation}

Furthermore, to obtain additional details on the sea surface (\emph{e.g.} foam positions, normal vectors...) the derivative of the horizontal and vertical displacements must be computed, respectively of $\mathbf{D}(\mathbf x,t)$ and $h(\mathbf x,t)$.
Taking advantage of the symmetry $\frac{\partial D_z}{\partial x}(\mathbf x,t) = \frac{\partial D_x}{\partial z}(\mathbf x,t)$, this relies on five derivatives only:

\begin{align*}
	\frac{\partial D_x}{\partial x}(\mathbf x,t) & = \sum_{\mathbf{k}} -ik_x \tilde{D}_x(\mathbf k,t)\exp\left(i\mathbf{k}\cdot\mathbf x\right),    \\
	\frac{\partial D_z}{\partial x}(\mathbf x,t) & = \sum_{\mathbf{k}} -ik_x \tilde{D}_z(\mathbf k,t)\exp\left( i \mathbf{k}\cdot \mathbf x\right), \\
	\frac{\partial D_z}{\partial z}(\mathbf x,t) & = \sum_{\mathbf{k}} -ik_z\tilde{D}_z(\mathbf k,t)\exp\left( i \mathbf{k}\cdot \mathbf x\right),  \\
	\frac{\partial h}{\partial x}(\mathbf x,t)   & = \sum_{\mathbf{k}} k \tilde{D}_x(\mathbf k,t)\exp\left( i \mathbf{k}\cdot \mathbf x\right),     \\
	\frac{\partial h}{\partial z}(\mathbf x,t)   & = \sum_{\mathbf{k}} k \tilde{D}_z(\mathbf k,t)\exp\left( i \mathbf{k}\cdot \mathbf x\right).
\end{align*}

\subsection{Cascades}
\label{subsec:cascades}
\begin{figure}
	\centering
	\tikzset{every picture/.style={line width=0.75pt}} 
	\begin{tikzpicture}
	\begin{axis}[smooth, enlargelimits=0.1, xmin=0.1,xmax=1.8,area legend,
			axis x line=bottom,
			axis y line=left,
			xlabel={Frequency ($Hz$)}, 
        	ylabel={Wave Spectra Density ($m^2/Hz$)}, 
        	]
		\addplot[pattern=north east lines,domain=0:0.5,pattern color =redF] coordinates
			{
				(0.39,0)(0.5,0.5)(0.7,3)
			}\closedcycle;
		\addplot[draw=none,forget plot] coordinates {
				(0.7, 3)
				(0.7, 0)
				(0, 0)  };

		\addplot[pattern=north east lines, domain=0.5:1.1, pattern color=bleuF] coordinates
			{
				(0.7,3)(0.8,5)(1,2)(1.1,1)
			}\closedcycle;
		\addplot[draw=none,forget plot] coordinates {
				(1.1, 0.7)
				(1.1, 0)
				(0.5, 0)  };

		\addplot[pattern=north east lines,domain=1.1:1.7,pattern color =jauneC] coordinates
			{
				(1.1,1)(1.4,0.1)(1.7,0)
			}\closedcycle;
		\addplot[draw=none,forget plot] coordinates {
				(2, 1)
				(2, 0)
				(1.1, 0)  };
		\legend{Cascade $0$,Cascade $1$,Cascade $2$}
	\end{axis}
\end{tikzpicture}
	\caption{Example of a spectrum split into $3$ different cascades: each cascade operates on a different interval of wave vector length.}
	\label{fig:cascades}
\end{figure}
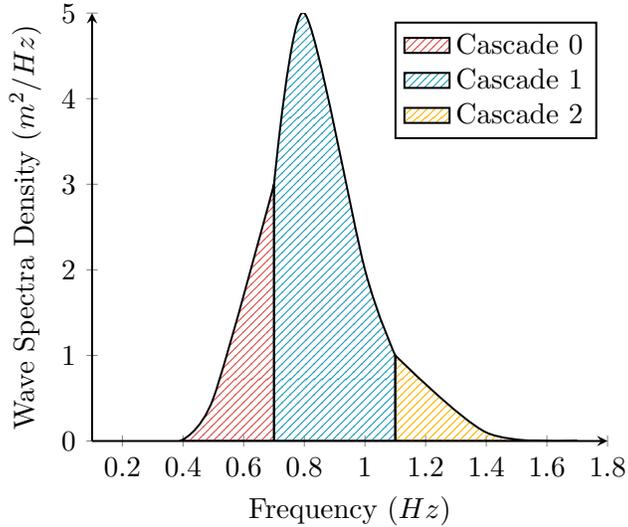

While the expression of the water height in Equation \ref{equa:heightWater} is convenient, it may result in an ocean surface that appears overly repetitive.
A naive solution to handle this problem consists in using a huge resolution $N$, but it has a major impact on performance.
\Framework uses the more efficient solution proposed by Dupuy and Bruneton (2012) \cite{dupuyRealtimeAnimationRendering2012}.
It consists in dividing the spectrum into $c$ layers or \emph{cascades} of different lengths.
Each of these layers represents a different wavelength in Equation~\ref{equa:heightWater}.
The length of the $i$-th cascade is denoted by $L_i$.
These lengths $L_i$ are parameters in the expression of the wave vector component described in Equation~\ref{equa:heightWater}.
Moreover, the different cascades are cut off at arbitrary value to avoid overlap between them, as illustrated in Figure~\ref{fig:cascades}.

\Framework uses by default $3$ cascades with the following settings:
\begin{itemize}
	\item $L_0 = 256$ operating for waves length $k\in [0,\frac{12\pi}{16}]$
	\item $L_1 = 16$ operating for waves length $k\in [\frac{12\pi}{16},\frac{12\pi}{4}]$
	\item $L_2 = 4$ operating for waves length $k\in [\frac{12\pi}{4},+\infty]$.
\end{itemize}

\subsection{Velocity of water}
\label{subsec:velWater}
Starting from the system \ref{equa:systemTessendorf}, \framework proposes to compute the velocity of the water at any depth $y$, as follows\footnote{See Appendix   \ref{sec:annexe} for more details on the origin of this expression}:
\begin{equation}\label{equa:velocity}
	\mathbf v(\mathbf x,y,t)=
	\sum_{\mathbf{k}} \tilde{\mathbf{v}}(\mathbf{k}, y, t)\exp\left(i\mathbf{k}\cdot\mathbf x\right),
\end{equation}
where the vector $\tilde{\mathbf{v}}(\mathbf{k}, y, t)$ is given as follows:
\begin{equation}
	\tilde{\mathbf{v}}\left(\mathbf{k}, y, t\right)=%
	E\left(\mathbf{k},y\right)\left(\tilde{h}_0\left(\mathbf{k}\right)\exp\left({i\omega(k)t}\right)-\tilde{h}_0^*\left(-\mathbf{k}\right)\exp\left({-i\omega(k)t}\right)\right)
	\left[
		\begin{array}{c}
			\displaystyle\frac{-k_xg}{\omega(k)} \\
			\displaystyle i\omega(k)             \\
			\displaystyle\frac{-k_zg}{\omega(k)}
		\end{array}
		\right],
\end{equation}
where the attenuation function $E(\mathbf{k},y)\in\mathbb R$ is defined as follows:
\begin{equation}
	\label{equa:attenuationDeep}
	E(\mathbf{k}, y)=
	\begin{cases}
		1+ky     & \text{if }y>0, \\
		\exp(ky) & \text{else}.
	\end{cases}
\end{equation}
The first case ($y>0$) comes from an extrapolation above the sea surface of the normal case ($y\leq0$).

To our knowledge, this expression of the velocity has never been used in a real time ocean simulator before.

\subsection{Waves' height implementation details}
This section gives some technical details of the implementation of an ocean with Tessendorf's method in \framework.

First, it should be noticed that the expressions of the ocean surface's height (Equation~\ref{equa:heightWater}), the displacements (Equation~\ref{equa:hori-displacement}) and their five derivatives, and the velocity (Equation~\ref{equa:velocity}) have “a form of an inverse discrete Fourier transform”.
More precisely, the \textbf{frequency space} coefficients appearing before the term $\exp(i\mathbf{k}\cdot\mathbf x)$ are stored in a square matrix of size $N\times N$, for each expression.
Then, the full sum is computed through an \textbf{inverse fast Fourier transform} (IFFT) applied on this coefficients' matrix.
In \framework, this trick allows computing these eight expressions without any significant impact on performance, even for huge domains.

Second, as \framework simulates a tile of water of size $N\times N$ using $c$ cascades, it stores the height, the displacement, and other relevant data in a 1D array of size $c$, where each entry contains a 2D array of size $N\times N$. \Framework computes efficiently these values on the GPU, with dedicated kernel that are launched with a grid of threads structured as $N \times N \times c$. Specifically, $N \times N$ threads are assigned to the $x$ and $y$ dimensions to cover the spatial resolution of the water tile, while $c$ threads along the $z$ dimension correspond to the different cascades. This approach ensures that computations for all cascades are executed in parallel.

Third, many terms are independent on the time $t$: $\tilde{h}_0(\mathbf{k})$, $\tilde{h}_0^*(-\mathbf{k})$ and the wave data ($\mathbf{k}$ and $\omega(k)$).
\Framework computes them for each cascade wave vector at each spectrum change only, but not at each iteration.
Then it does a combination with time-dependent terms at each iteration, giving an expression for the coefficients in frequency space for the eight IFFT.

To compute the water height at a specific point, it is not sufficient to directly evaluate Equation \ref{equa:heightWater} after performing the IFFT. This is because the water surface is described parametrically, with fluid positions depending on both horizontal and vertical displacements, as detailed in Section \ref{subsec:water_displacement_and_derivative}. Accurately determining the water surface position at a given horizontal location requires an iterative process (see Algorithm \ref{alg:compute_water_height}): the estimated position is adjusted by subtracting the displacement vector, and the water height is recalculated until convergence. However, in practice, waiting for full convergence may be computationally expensive. Instead, the loop is limited to $N_{iter}=4$ iterations, a value chosen to balance accuracy and performance (see Appendix \ref{sec:evaluation_of_the_number_of_iterations_for_convergence} for details on this choice).

\begin{algorithm}
\caption{Iterative Retrieval of Water Height}
\label{alg:compute_water_height}
\begin{algorithmic}[1]
\Require A position $\mathbf x$ to evaluate the water height and the horizontal and vertical displacements: $\mathbf{D}(\mathbf x,t)$, $h(\mathbf x,t)$
\State Initialize displacement $\mathbf{W} = (0,0,0)$
\For{$k = 1$ to $N_{\text{iter}}$} \Comment{Iterative correction to account for displacement}
    \State Compute shifted position: $\mathbf{W} = \mathbf{x} - \mathbf{W}$
    \State Compute horizontal displacement at $\mathbf{W}$ 
    \Comment{The sum of bilinear interpolation of the displacement on each cascades}
\EndFor
\State \Return The vertical component of $W$
\end{algorithmic}
\end{algorithm}

Last but not least, one might think that applying IFFT should be enough to get the expected result.
Nevertheless, using the standard IFFT algorithm (such as the one of Cooley and Tukey (1965) \cite{cooleyAlgorithmMachineCalculation1965}) leads to some incorrect geometry changes.
Unlike usual IFFT, the used range is $\left[-N/2, N/2\right]$ and so the wave vectors have negative components (\emph{e.g.} the height).
To fit with the theory of Fourier transforms, in \framework each value at index $i,j$ of the 2D matrices is multiplied by $(-1)^{i+j}$.

In \framework, the number of IFFT applications is divided by two using Theorem \ref{theorem:ifft} on Hermitian matrices\footnote{A matrix $X=(x_{n,m})$ is Hermitian if it is equal to its own conjugate transpose: $X = X^{\dagger}$ or $x_{n,m}=x^*_{m,n}$.} (see annex \ref{sec:annexe-2IFFT} for its proof).
\begin{theorem}
	\label{theorem:ifft}
	Let: $X=(x_{n,m}), Y=(y_{n,m}) \in M_n(\mathbb{C})$ be two Hermitian matrices.
	Then the following relation is valid:
	\begin{equation}
		\mathcal{F}^{-1}\left(X+iY\right) =
		\Re\left(\mathcal{F}^{-1}\left(X\right)\right)
		+ i\Re\left(\mathcal{F}^{-1}\left(Y\right)\right)
	\end{equation}
\end{theorem}
Hence, in \framework the coefficients in frequency space of two Hermitian matrices $X$ and $Y$ are combined in a single matrix $H=X+iY$.
Using Theorem \ref{theorem:ifft}, one complex IFFT suffices to obtain the real parts of two IFFT $\mathcal{F}^{-1}\left(X\right)$ and $\mathcal{F}^{-1}\left(Y\right)$.
This treatment allows reducing IFFT computation times by a factor close to 2.

\subsection{Waves' velocity implementation details}
Contrary to the displacement and its derivative which are height maps and 2D functions, the velocity must be known in 3D space.
Hence, its calculation is more difficult and requires more computation time.
Computing the velocity for all the points of a 3D grid, or even only the points where it must be evaluated for fluid to solid coupling (Section~\ref{subsubsec:fluidToSolidCoupling}) is not possible in real-time, as it involves a large to huge number of IFFT.
Unfortunately, the dependency of the attenuation (Equation \ref{equa:attenuationDeep}) on both the wave vectors and $y$ makes the depth dependency of the velocity fundamentally tied to the wave vector, preventing any simplifications that could reduce the required calculations.

To reduce the complexity of this problem, \framework opts for an interpolation scheme, involving a reasonable number of IFFTs that is proportional to the size of the discretization.

More concretely, the velocity must be computed at any horizontal position in $I=[y_{min}, y_{max}]$.
In \framework, $d$ growing depths $y_i$ are chosen in this interval $I$.
For each of these depths $y_i$, the $i$-th velocity $\mathbf{v_i}(\mathbf x,y_i,t)$ is computed using one IFFT, leading to $d$ IFFTs for all the sampled depths $y_i$.
Then, the velocity $\mathbf{v}(\mathbf x,y,t)$ at any depth $y\in I$ is computed using an ad hoc interpolation from the different $\mathbf{v_i}$.

A first way to do this interpolation mechanism relies on a linear scheme.
Nevertheless, it produces a very inaccurate result.
Hence, \framework uses an exponential scheme made on logarithmic sampling of the interval $I$.
They are discussed in the next two sections.

\subsubsection{Exponential interpolation}
The vertical dependency of velocity links to the exponential attenuation function from Equation \ref{equa:attenuationDeep}.
Hence, instead of a linear interpolation \framework uses an exponential interpolation $f_e(x)$ that should resemble the following expression:
\begin{equation}
	f_e(x)=\alpha \exp(\beta x).
\end{equation}
Knowing $f_e(a)=f(a)$ and $f_e(b)=f(b)$, it is straightforward to obtain the two values $\alpha$ and $\beta$ as follows:
\begin{equation}
	\begin{cases}
		\displaystyle\alpha = \frac{f(a)}{\exp(\beta a)} \\
		\displaystyle\beta = \frac{\ln(|f(b)|) - \ln(|f(a)|)}{b - a}
	\end{cases}
\end{equation}
This exponential interpolation function $f_e(x)$ is applied to extract the magnitude and rotation angle of the velocity $\mathbf v(\mathbf x,y,t)$.

\subsubsection{Logarithmic distribution}
As the velocity decays exponentially with depth, the ones close to the ocean surface play a more important role in accuracy than the one at the bottom of the ocean.
Hence, a regular sampling along the depth seems inappropriate, albeit the exponential interpolation.

\Framework uses the following logarithmic distribution:
\begin{equation}
	l_d(y_i) =
	\begin{cases}
		\beta \ln(\alpha|y_i|^2+1) ~~\text{if}~y_i>0     \\
		-\beta \ln(\alpha|y_i|^2+1) ~~\text{if}~y_i\leq0 \\
	\end{cases}
\end{equation}
where $\alpha=0.0001$ and $\beta = -y_{min}/\left(2 \ln(\alpha |y_{min}|^2 + 1)\right)$.

This distribution is designed based on the following considerations.
First, while the velocity decays exponentially, it relies on a logarithmic function similar to $y_i\mapsto \ln(|y_i|+1)$.
Since this function transforms the negative $y_i$ into a positive one, the result is multiplied by the sign of $y_i$.
Also, this function decreases slowly (actually with $\ln(250)=5.5$), hence the logarithmic distribution uses the square of $|y_1|$.
Lastly, the parameter $\alpha$ controls the “squeezing” in the center, while the parameter $\beta$ controls the deepest value of $l_d$.
Notice that $\beta$ is chosen as the solution of $l_d(y_{min})=\frac{y_{min}}{2}$ but not at $y_{min}$: it is the solution giving the best accuracy after multiple tries.

\begin{figure}[ht]
	\centering
	\subfloat{\includegraphics[width=0.7\textwidth]{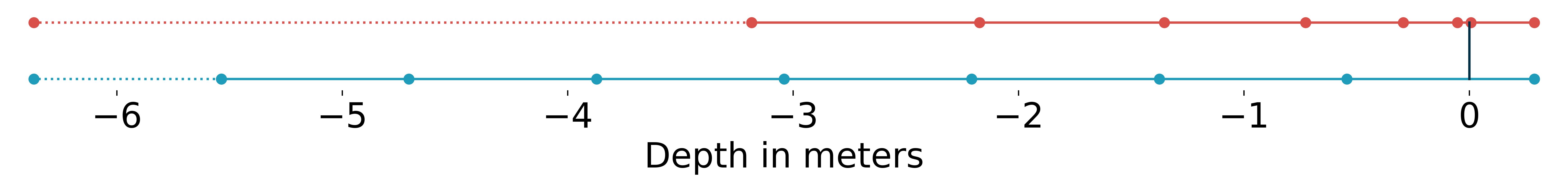}\label{fig:gridInterpolationDegee8}}
	\hfill
	\subfloat{\includegraphics[width=0.3\textwidth]{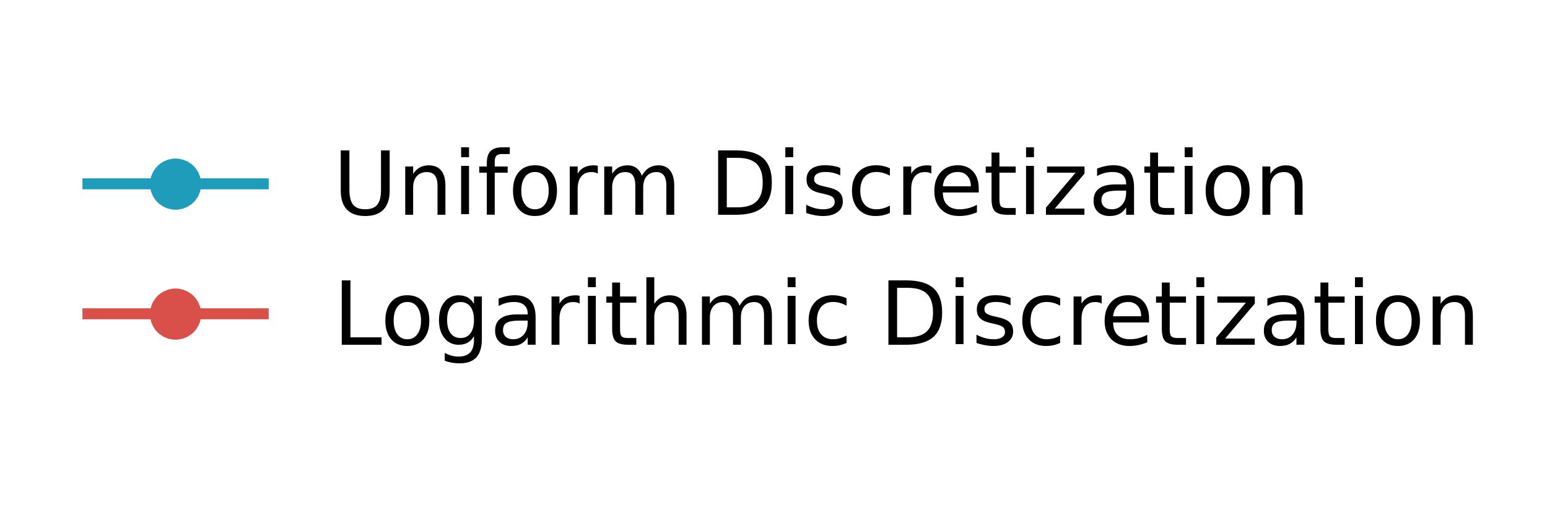}\label{fig:gridInterpolationDegree8Legend}}
	\caption{This figure shows a comparison between a \textcolor{bleuF}{uniform discretization in blue} and a \textcolor{redF}{logarithmic discretization in red}. The dotted parts represent the last interpolation toward a null velocity at $y_{min}$.}
	\label{fig:comparisonGrid}
\end{figure}

Figure \ref{fig:comparisonGrid} compares the logarithmic discretization to the uniform discretization. As expected, the behavior with the former leads to fewer points close to $y_{min}$ and more points close to $y_{max}$.
Note also the higher density of points close to $y=0$.

Moreover, using the logarithmic distribution leads to $l_d(y_{min})=\frac{y_{min}}{2}$.
Assuming that the velocity at $y_{min}$ is null, then for any depth between $y_{min}$ and $\frac{y_{min}}{2}$ the velocity can be computed between the one at $\frac{y_{min}}{2}$ and $0$.

\begin{figure}[ht]
	\centering
	\includegraphics[width=0.8\textwidth]{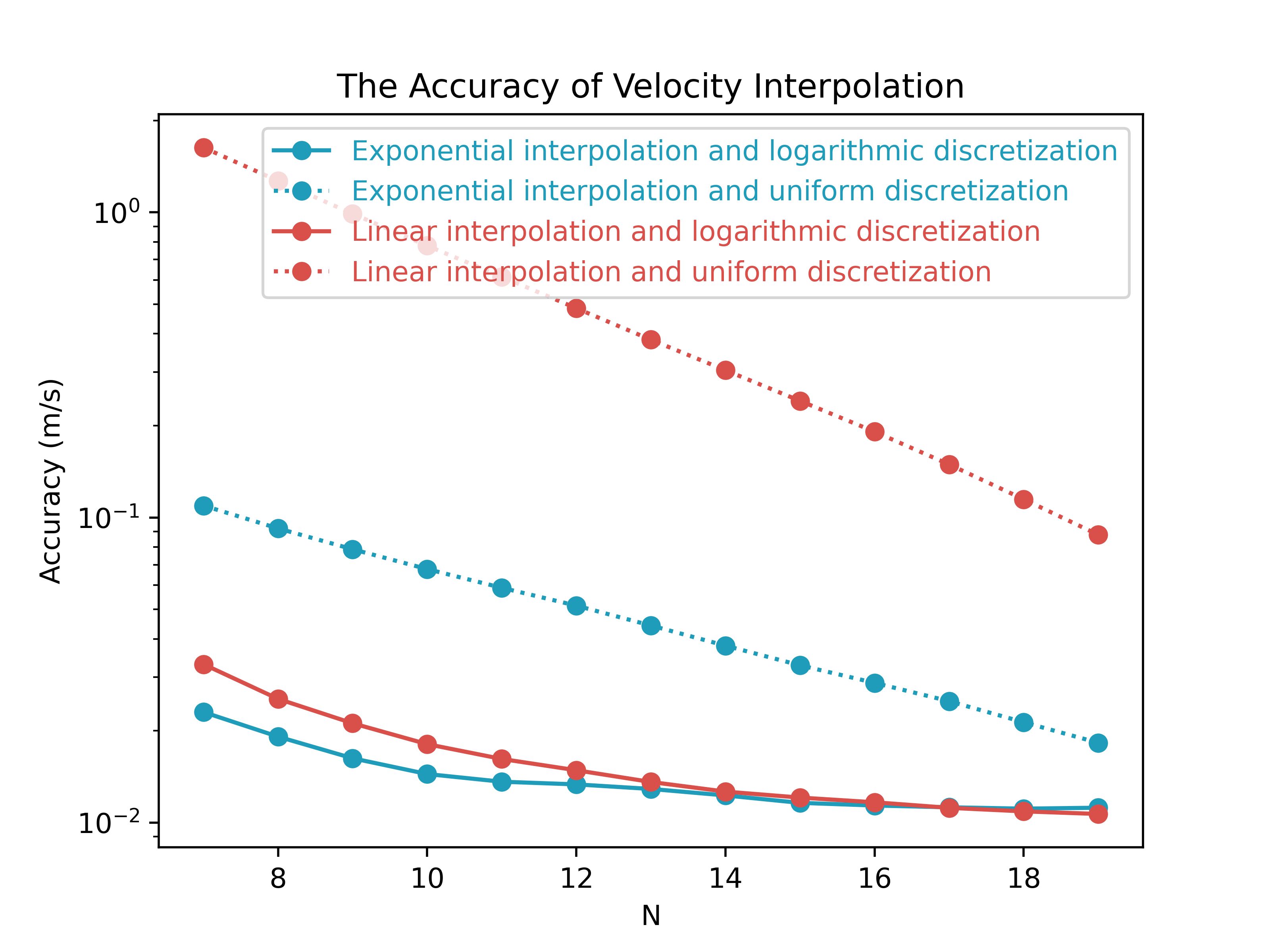}\label{subfig:interpolationAccuracyComparisonLogScale}
	\caption{This figure represents the accuracy of different type of interpolation in function of the size of the discretization $N$. The accuracy axis used a logarithmic scale to be more readable.
		It uses 10000 positions from a continuous uniform distribution in the parallelepiped $[-1000,1000]\times[-125,4.5]\times[-1000, 1000]$ and with a wind speed of 20 $m.s^{-1}$.
		The interpolated velocity $\tilde{\mathbf{v}}$ and the analytical velocity $\mathbf{v}$ are evaluated.
		The accuracy is defined by the mean of all positions of $\big|||\tilde{\mathbf{v}}||-||\mathbf{v}||\big|$.
		It can be noticed that considering a uniform discretization, then exponential interpolation has a better accuracy than linear one.
		Logarithmic discretization is much better than uniform discretization, whatever the used interpolation.
		At least, for low $N$, exponential interpolation gives slightly better result than linear interpolation.}
	\label{fig:interpolationAccuracyComparison}
\end{figure}

Figure \ref{fig:interpolationAccuracyComparison} compares the accuracy using linear mechanism and the exponential interpolation with the logarithmic discretization.
Clearly the latter gives much better accuracy, as expected.

\subsubsection{Choosing the interpolation degree}

Knowing the interpolation method, the interpolation degree $d$ (number of points in the table) must be decided.
Note, that using more points results in better accuracy, but also more computation time.
Then, a trade-off between performance and accuracy must be determined.
With this aim, the \textit{multi-objective optimization} method is used.

Briefly, this method reduces a multi-criteria optimization problem into a one dimensional one.
It defines a one dimensional function as a linear sum of different objective functions; here, it concerns the performance and accuracy for a given interpolation degree.

Hence, this optimization function $J$ is defined as follows:
\begin{equation}
	J(d) = \alpha P(d) + \beta A(d)
\end{equation}
where $d$ is the interpolation degree, $P(d)$ measures the performance of the velocity interpolation, $A(d)$ measures the accuracy of the velocity interpolation, and $\alpha$ and $\beta$ are the weight arbitrary assigned to the performances and the accuracy functions.

\begin{figure}[ht]
	\centering
	\includegraphics[width=0.8\textwidth]{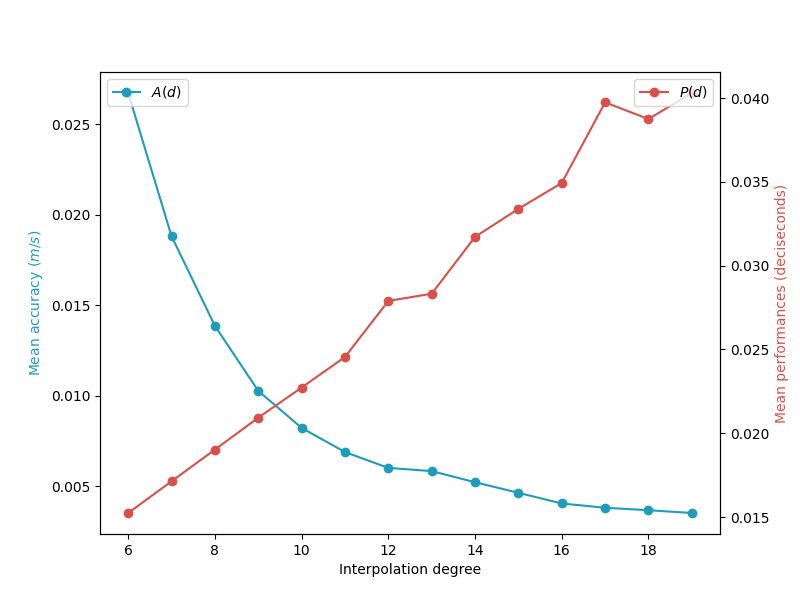}
	\caption{These figures represent the different objective for choosing the interpolation degree. In blue, the mean accuracy of velocity interpolation in meters per seconds, we denote this function $A(d)$. In red, the mean performances of velocity interpolation in deciseconds, we denote this function $P(d)$. We estimate this function with $10000$ sample.}
	\label{fig:interpolationCriteria}
\end{figure}

To make $P(d)$ and $A(d)$ having the same order of magnitude, the units are the \textbf{seconds} for performance and the \textbf{meter per seconds} for accuracy (see figure \ref{fig:interpolationCriteria}).
Moreover, the performance being the more important for real-time process, the following arbitrary choice are made for the coefficients of $J$: $\alpha = 10$ and $\beta = 2$.

\begin{figure}[ht]
	\centering
	\includegraphics[width=0.8\textwidth]{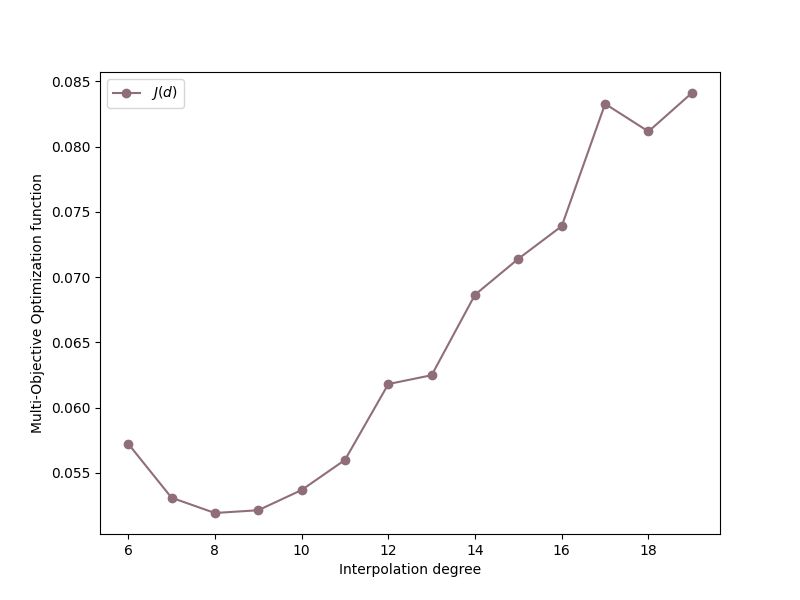}
	\caption{Multi-objective function $P$ to optimize, with two objectives: performance and accuracy.
		Its minimum is reached by the interpolation degree of $8$. For more details on each objective, see Figures \ref{fig:interpolationCriteria} which represent them individually.}
	\label{fig:multiObjectiveFunction}
\end{figure}

Figure \ref{fig:multiObjectiveFunction} shows the function $J$ using these parameters. It appears that its minimum is obtained with $d=8$.
It is the value used in \framework.

\subsection{Coupling solid and fluid}
\label{sec:couplingSolidAndFluid}
The physical simulation based on the Tessendorf's method is very fast, and the spectrum approach leaves a lot of control to the user.
However, it doesn't simulate the interaction between the fluid and a solid, which is needed for adding some boats in the simulation.
To overcome this flaw, \framework uses two methods: a first one for fluid-to-solid action, and a second one for solid-to-fluid action.

These two methods are designed with the following constraints in mind:
\begin{enumerate}
	\item \label{enum:simpleParam} Rely on simple parameterization for non-physicists, to be easily usable by artists;
	\item \label{enum:noPrecal} Not requiring too many pre-calculations, to avoid an overly complex framework;
	\item \label{enum:perf} Have excellent performance (a few milliseconds at most), so that \framework reaches real time globally, including all simulation and rendering stages.
\end{enumerate}

\section{Fluid-to-solid action}
\label{sec:fluidToSolidAction}
The action of the fluid on a given solid can be approximated by a \emph{set of forces} acting continuously on the \emph{contact surface} between the fluid and the solid.
In \framework, the contact surface is represented using the triangulation of the solid's mesh, leading to a discrete approximation of the set of forces.
Moreover, to remain real-time \framework uses forces computed from simple geometric and physical parameters only.
To maintain realistic physics after these discrete approximations and those described in subsequent sections, it is important to use a closed mesh, uniformly triangulated. In addition, for meshes representing vessels, ensuring symmetry is crucial to maintain stability.

\subsection{Geometrical Parameters}
\label{subsec:geomParam}
In \framework, the forces are computed for each boat or body, independently of the others.
The calculation of the forces needs the following geometrical data from a body and the water surface:
the lower hull (part of the hull in the water), the waterline, and the upper hull (part of the hull above the water).

At first, \framework checks the vertical position of body's each vertex relative to the water surface.
Then, it assigns a status to each body's face: submerged, partially submerged or not submerged.
The triangles that are partially submerged are divided as done in Kerner (2016) \cite{kernerWaterInteractionModel2016}, producing the waterline as the intersection edges between the mesh and the water surface.
This waterline is used later to compute the mask in Section~\ref{subsec:mask}.
Then the body surface is decomposed into a set $\left\lbrace T_i^s\right\rbrace$ of submerged triangles, and $\left\lbrace T_i^{\neg s}\right\rbrace$ of not submerged ones.
\Framework computes the area of each submerged or not submerged triangle at this step too.

The buoyancy calculation presented in Section \ref{subsubsec:buoyancy} requires the full submerged volume $v_w$ of the mesh.
\Framework uses the prism approximation similarly to the approximation proposed by Bajo \etal (2020) \cite{bajoRealisticBuoyancyModel2020}.
While Bajo \etal compute the submerged volume at the texel level, the same principle can be applied using triangles instead of texels using as they did the divergence theorem.
Specifically, $v_w$ is approximated as the sum of the prisms with bases corresponding to the triangle areas $A_i$ and heights defined by the distance $d_i$ between the triangle centers $p_i$ and the water surface $h(p_i,t)$.
Each term in the sum is weighted by the vertical component of the triangle normal $n_i^y$:
\begin{equation}
	v_w = \sum_{T_i^s} A_i d_i n_i^y.
\end{equation}
Note that this formula is correct only under the assumption that the mesh is closed.

The pseudocode for the all geometrical calculations performed in this section can be read in the Algorithm \ref{alg:geom_params}.

\begin{algorithm}
\caption{Computation of Geometrical Parameters}
\label{alg:geom_params}
\begin{algorithmic}[1]
\Require List of triangles $T_i$ with vertex positions $\mathbf{v}_i$ and normal $\mathbf{n}_i$

\For{each triangle $T_i = (\mathbf{v}_0, \mathbf{v}_1, \mathbf{v}_2)$ with normal $\mathbf{n}_i$}
    \State Retrieve water height $d_0$, $d_1$, and $d_2$ at vertices $\mathbf{v}_0$, $\mathbf{v}_1$, and $\mathbf{v}_2$
    \State Compute area: $A_i = \frac{1}{2} \|(\mathbf{v}_1 - \mathbf{v}_0) \times (\mathbf{v}_2 - \mathbf{v}_0)\|$
    \State Compute centroid: $\mathbf{p}_i = \frac{1}{3} (\mathbf{v}_0 + \mathbf{v}_1 + \mathbf{v}_2)$
    \State Compute centroid depth: $d_i = \frac{1}{3} (d_0 + d_1 + d_2)$
    
    \If{$d_0 \geq 0$ and $d_1 \geq 0$ and $d_2 \geq 0$}  \Comment{Fully above water}
        \State Set submerged area $A_i^s = 0$
        \State Set volume contribution $v_{w_i} = 0$
        
    \ElsIf{$d_0 < 0$ and $d_1 < 0$ and $d_2 < 0$} \Comment{Fully submerged}
        \State Set submerged area $A_i^s = A_i$
        \State Compute submerged volume: $v_{w_i} = A_i d_i n_i^y$
        
    \ElsIf{one vertex is above the water surface (e.g., $d_0 \geq 0$)} \Comment{Partially submerged}
        \State Compute intersection points $\mathbf{p}_1, \mathbf{p}_2$ using linear interpolation:
        \State $\mathbf{p}_j = \mathbf{v}_j + \alpha (\mathbf{v}_k - \mathbf{v}_j)$ where $\alpha = \frac{d_j}{d_j - d_k}$
        \State Compute submerged area $A_i^s$ from new submerged sub-triangle(s)
        \State Compute submerged volume: $v_{w_i} = A_i^s d_i n_i^y$
        
    \ElsIf{two vertices are above the water surface (e.g., $d_0 \geq 0$, $d_1 \geq 0$)} \Comment{Mostly submerged}
        \State Compute intersection points $\mathbf{p}_1, \mathbf{p}_2$ using linear interpolation
        \State Compute submerged area $A_i^s$ from the remaining submerged sub-triangle
        \State Compute submerged volume: $v_{w_i} = A_i^s d_i n_i^y$
    \EndIf
\EndFor

\end{algorithmic}
\end{algorithm}

At last, from all the submerged faces $T_i^s$ \framework computes the center of immersion $c_i$ used for the buoyancy calculation as presented in Section~\ref{subsec:forces}.

\subsection{Forces}
\label{subsec:forces}
\Framework uses three forces as in Kellomaki (2014) \cite{kellomakiRigidBodyInteraction2014}:
\begin{enumerate}
	\item Buoyancy $\mathbf{F}_b$;
	\item Water drag $\mathbf{F}_w$;
	\item Air drag $\mathbf{F}_a$.
\end{enumerate}
These forces are computed for each triangle of the lower or upper hull, and so after the different geometric parameters being calculated.

\subsubsection{Buoyancy}
\label{subsubsec:buoyancy}
The buoyancy is computed once for the whole boat from the following analytical formula:
\begin{equation}
	\mathbf{F_b} = - v_w \rho_w \mathbf g,
\end{equation}
where $v_w$ is the submerged volume in $m^3$ (cf. Section~\ref{subsec:geomParam}) and $\rho_w$ is the water density.
The water density is obtained at any depth by linear interpolation using the measure of the International Towing Tank Conference (ITTC) \cite{ittcFreshWaterSeawater2011}.
The buoyancy is applied to the center of immersion $c_i$.

\subsubsection{Drag Forces}
\label{subsubsec:dragForce}
\Framework calculates drag forces based on the relative velocity between the fluid medium (water or air) and the submerged or exposed portion of the body. This approach is inspired by previous works \cite{yukselRealtimeWaterWaves2010,kellomakiRigidBodyInteraction2014,kernerWaterInteractionModel2016}.

The drag force for each triangle $T_i$ of the body is computed as:
\begin{equation}
	\mathbf{F_d}= -\frac{1}{2} C_d \rho A_i^{\perp} \left\Vert\mathbf{v_{i_{rel}}}\right\Vert\mathbf{v_{i_{rel}}},
\end{equation}
where:
\begin{itemize}
	\item $\mathbf{v_{i_{rel}}}=\mathbf{v_i}-\mathbf{v_{i}^{m}}$ is the relative velocity in $m.s^{-1}$ of the triangle $T_i$, equal to the difference between its velocity $\mathbf{v_i}$ and the velocity $\mathbf{v_{i}^{m}}$ of the fluid medium (water or air) at its center;
	\item $A_i^{\perp}$ is the projected area in $m^2$ of the triangle $T_i$ with respect to the relative velocity unit direction (\ie the vector $\mathbf{v_{i_{rel}}}/\left\Vert\mathbf{v_{i_{rel}}}\right\Vert$);
	\item $\rho$ is the density of the fluid medium in $kg.m^{-3}$ ($\rho_w$ as defined in Section~\ref{subsubsec:buoyancy} for water and $1.204~kg.m^{-3}$  for air\footnote{According to the international standard atmosphere, the air density at sea level for $20$°C is $1.204~kg.m^{-3}$ cf. \url{https://www.digitaldutch.com/atmoscalc/table.htm}});
	\item $C_d$ is the drag coefficient of the body in the fluid medium, specified by the user.
\end{itemize}

For the submerged part $\left\lbrace T_i^s\right\rbrace$, the medium is water, while for the exposed part $\left\lbrace T_i^{\neg s}\right\rbrace$, the medium is air.

\subsection{Fluid-to-solid algorithm}
\label{subsec:algoForces}

To ensure a structured and efficient data workflow on the GPU, all computations for a given mesh are performed separately, avoiding data interleaving between different meshes. Each kernel invocation processes a single mesh, iterating over its triangles before moving to the next mesh.

Firstly, for each triangle in the mesh, the geometric parameters are calculated and then used to compute the drag forces. These include necessary attributes such as the submerged volume, normal vectors, and the drag forces acting on submerged and non-submerged parts.

Once per-triangle computations are completed, the forces and geometrical parameters must be integrated into the physics solver, which operates on the CPU. To minimize data transfers and reduce latency, a parallel reduction kernel is applied on the GPU for each mesh individually. This step aggregates all per-triangle forces and parameters into global values for the entire mesh.

Buoyancy, however, is handled differently. Since it depends solely on the total submerged volume rather than individual triangles, it is computed separately using an analytical formula and applied directly at the submerged center without requiring per-triangle summation.

With all global parameters computed, the final step transfers the data to the CPU, where the physics solver integrates the buoyancy and the water drag forces are applied on the submerged center, while air drag forces are applied at the center of the non-submerged part.

This pipeline is presented in Algorithm \ref{alg:fluid_to_solid}.

\begin{algorithm}
\caption{Fluid-to-Solid Interaction Algorithm}
\label{alg:fluid_to_solid}
\begin{algorithmic}[1]
\For{each mesh} \Comment{Iterate over all mesh}
    \For{all triangles $T_i$} \Comment{Iterate over all triangles}
        \State Compute geometrical parameters \Comment{Algorithm \ref{alg:geom_params}}
    \EndFor
\EndFor

\For{each mesh} \Comment{Iterate over all mesh}
    \For{all triangles $T_i$} \Comment{Iterate over all triangles}
        \State Compute drag forces $\mathbf{F_d}$ for each triangles\Comment{Section \ref{subsubsec:dragForce}}
    \EndFor
\EndFor

\For{each mesh} \Comment{Iterate over all mesh}
    \For{all triangles $T_i$} \Comment{Iterate over all triangles}
        \State Apply parallel reduction to aggregate forces and geometrical parameters.
        \EndFor
\EndFor

\For{each mesh} \Comment{Iterate over all mesh}
    \State Transfer reduced forces and parameters to the CPU
    \State Compute buoyancy force $\mathbf{F_b}$ from geometrical parameters \Comment{Section \ref{subsubsec:buoyancy}}
    \State Integrate total forces and moments in the physics engine
\EndFor
\end{algorithmic}
\end{algorithm}

\begin{figure}[!h]
	\centering
\begin{tikzpicture}[x=0.75pt,y=0.75pt,yscale=-1,xscale=1]

\draw  [color=noirC  ,draw opacity=1 ] [line join = round][line cap = round] (323.05,134.3) .. controls (321.49,101.18) and (319.93,68.07) .. (318.38,34.96) ;
\draw  [color=noirC  ,draw opacity=1 ][fill=noirC  ,fill opacity=1 ] (298.53,43.21) -- (319.45,49.25) -- (319.05,35.88) -- cycle ;
\draw [color=jauneC  ,draw opacity=1 ][line width=1.5]   (311.73,180.27) -- (311.73,93.27) ;
\draw [shift={(311.73,90.27)}, rotate = 90] [fill=jauneC  ,fill opacity=1 ][line width=0.08]  [draw opacity=0] (8.93,-4.29) -- (0,0) -- (8.93,4.29) -- cycle    ;

\draw  [color=bleuF  ,draw opacity=1 ][line width=1.25] (124,169.13) .. controls (128.08,169.45) and (131.98,169.76) .. (136.5,169.76) .. controls (141.02,169.76) and (144.92,169.45) .. (149,169.13) .. controls (153.08,168.81) and (156.98,168.5) .. (161.5,168.5) .. controls (166.02,168.5) and (169.92,168.81) .. (174,169.13) .. controls (178.08,169.45) and (181.98,169.76) .. (186.5,169.76) .. controls (191.02,169.76) and (194.92,169.45) .. (199,169.13) .. controls (203.08,168.81) and (206.98,168.5) .. (211.5,168.5) .. controls (216.02,168.5) and (219.92,168.81) .. (224,169.13) .. controls (228.08,169.45) and (231.98,169.76) .. (236.5,169.76) .. controls (241.02,169.76) and (244.92,169.45) .. (249,169.13) .. controls (253.08,168.81) and (256.98,168.5) .. (261.5,168.5) .. controls (266.02,168.5) and (269.92,168.81) .. (274,169.13) .. controls (278.08,169.45) and (281.98,169.76) .. (286.5,169.76) .. controls (291.02,169.76) and (294.92,169.45) .. (299,169.13) .. controls (303.08,168.81) and (306.98,168.5) .. (311.5,168.5) .. controls (316.02,168.5) and (319.92,168.81) .. (324,169.13) .. controls (328.08,169.45) and (331.98,169.76) .. (336.5,169.76) .. controls (341.02,169.76) and (344.92,169.45) .. (349,169.13) .. controls (353.08,168.81) and (356.98,168.5) .. (361.5,168.5) .. controls (366.02,168.5) and (369.92,168.81) .. (374,169.13) .. controls (378.08,169.45) and (381.98,169.76) .. (386.5,169.76) .. controls (391.02,169.76) and (394.92,169.45) .. (399,169.13) .. controls (403.08,168.81) and (406.98,168.5) .. (411.5,168.5) .. controls (416.02,168.5) and (419.92,168.81) .. (424,169.13) .. controls (428.08,169.45) and (431.98,169.76) .. (436.5,169.76) .. controls (441.02,169.76) and (444.92,169.45) .. (449,169.13) .. controls (453.08,168.81) and (456.98,168.5) .. (461.5,168.5) .. controls (466.02,168.5) and (469.92,168.81) .. (474,169.13) .. controls (478.08,169.45) and (481.98,169.76) .. (486.5,169.76) .. controls (491.02,169.76) and (494.92,169.45) .. (499,169.13) .. controls (503.08,168.81) and (506.98,168.5) .. (511.5,168.5) .. controls (516.02,168.5) and (519.92,168.81) .. (524,169.13) .. controls (528.08,169.45) and (531.98,169.76) .. (536.5,169.76) .. controls (541.02,169.76) and (544.92,169.45) .. (549,169.13) .. controls (549.52,169.09) and (550.04,169.05) .. (550.56,169.01) ;
\draw   (434.66,131.22) .. controls (434.66,131.36) and (434.67,131.49) .. (434.67,131.63) .. controls (435.82,164.74) and (386.65,193.32) .. (324.86,195.45) .. controls (263.07,197.59) and (212.05,172.48) .. (210.9,139.37) .. controls (210.88,138.77) and (210.88,138.17) .. (210.89,137.57) -- cycle ;
\draw  [color=jauneC  ,draw opacity=1 ][line width=1.2] (307.43,177.01) -- (315.31,184.5)(314.93,177.01) -- (307.81,184.5) ;
\draw  [color=redF  ,draw opacity=1 ][line width=1.2] (326.18,143.79) -- (334.06,151.28)(333.68,143.79) -- (326.56,151.28) ;
\draw  [color=greenF  ,draw opacity=1 ][line width=1.2] (317.18,159.76) -- (325.06,167.25)(324.68,159.76) -- (317.56,167.25) ;
\draw [color=noirC  ,draw opacity=1 ] [line width=1.5]  [dash pattern={on 0.84pt off 2.51pt}](380.73,75.93) -- (460.73,45.4) ;
\draw [shift={(463.53,44.33)}, rotate = 159.11] [fill=noirC  ,fill opacity=1 ][line width=0.08]  [draw opacity=0] (8.93,-4.29) -- (0,0) -- (8.93,4.29) -- cycle    ;
\draw [color=redF  ,draw opacity=1 ][line width=1.5]   [dash pattern={on 0.84pt off 2.51pt}](380.73,75.93) -- (342.16,52.44) ;
\draw [shift={(339.6,50.88)}, rotate = 31] [fill=redF  ,fill opacity=1 ][line width=0.08]  [draw opacity=0] (8.93,-4.29) -- (0,0) -- (8.93,4.29) -- cycle    ;
\draw [color=bleuF  ,draw opacity=1 ][line width=1.5]   [dash pattern={on 0.84pt off 2.51pt}](380.73,75.93) -- (409.73,75.9) ;
\draw [shift={(412.73,75.9)}, rotate = 179.94] [fill=bleuF  ,fill opacity=1 ][line width=0.08]  [draw opacity=0] (8.93,-4.29) -- (0,0) -- (8.93,4.29) -- cycle    ;
\draw [color=greenF  ,draw opacity=1 ][line width=1.5]   (321,163.33) -- (321,235.33) ;
\draw [shift={(321,238.33)}, rotate = 270] [fill=greenF  ,fill opacity=1 ][line width=0.08]  [draw opacity=0] (8.93,-4.29) -- (0,0) -- (8.93,4.29) -- cycle    ;
\draw [color=purpleF  ,draw opacity=1 ][line width=1.5]   (311.27,180.63) -- (361.16,150.9) ;
\draw [shift={(363.73,149.37)}, rotate = 149.21] [fill=purpleF  ,fill opacity=1 ][line width=0.08]  [draw opacity=0] (8.93,-4.29) -- (0,0) -- (8.93,4.29) -- cycle    ;
\draw [color=redF  ,draw opacity=1 ][line width=1.5]   (330.07,147.2) -- (301.23,159.14) ;
\draw [shift={(298.46,160.29)}, rotate = 337.51] [fill=redF  ,fill opacity=1 ][line width=0.08]  [draw opacity=0] (8.93,-4.29) -- (0,0) -- (8.93,4.29) -- cycle    ;

\draw (298.12,99.03) node   [align=left] {\begin{minipage}[lt]{15.46pt}\setlength\topsep{0pt}
$\displaystyle \mathbf{\textcolor{jauneC}{F}}\textcolor{jauneC}{_{b}}$
\end{minipage}};
\draw (332.37,225.08) node   [align=left] {\begin{minipage}[lt]{15.46pt}\setlength\topsep{0pt}
$\displaystyle \mathbf{\textcolor{greenF}{F}}\textcolor{greenF}{_{g}}$
\end{minipage}};
\draw (379.74,149) node   [align=left] {\begin{minipage}[lt]{15.46pt}\setlength\topsep{0pt}
$\displaystyle \mathbf{\textcolor{purpleF}{F}}\textcolor{purpleF}{_{w}}$
\end{minipage}};
\draw (297.07,144.96) node   [align=left] {\begin{minipage}[lt]{15.46pt}\setlength\topsep{0pt}
$\displaystyle \mathbf{\textcolor{redF}{F}}\textcolor{redF}{_{a}}$
\end{minipage}};
\draw (423.23,48.8) node   [align=left] {\begin{minipage}[lt]{16.65pt}\setlength\topsep{0pt}
$\displaystyle \mathbf{\textcolor{noirC}{v}}$
\end{minipage}};
\draw (356.74,88.41) node  [color=redF  ,opacity=1 ] [align=left] {\begin{minipage}[lt]{15.46pt}\setlength\topsep{0pt}
$\displaystyle \mathbf{v}_{a}$
\end{minipage}};
\draw (394.5,88.07) node  [color=bleuF  ,opacity=1 ] [align=left] {\begin{minipage}[lt]{15.46pt}\setlength\topsep{0pt}
$\displaystyle \mathbf{v}_{w}$
\end{minipage}};

\end{tikzpicture}
	\caption{Summary of forces calculation. Using the same notation as Section~\ref{subsec:forces} it follows:
		the buoyancy force \textcolor{jauneC}{$F_b$} in \textcolor{jauneC}{yellow}, applied to the center of immersion;
		the sum of water drag force \textcolor{purpleF}{$F_w$} in \textcolor{purpleF}{purple}, applied to the center of immersion too;
		the sum of the air drag forces \textcolor{redF}{$F_a$} in \textcolor{redF}{red}, applied to the center of the non-submerged part;
		and finally the gravity force \textcolor{greenF}{$F_g$} in \textcolor{greenF}{green}, applied to the center of gravity.
		Different velocities are displayed above the ship:
		the wind velocity \textcolor{redF}{$v_a$} in \textcolor{redF}{red};
		the water velocity \textcolor{bleuF}{$v_w$} in \textcolor{bleuF}{blue};
		and the object velocity \textcolor{noirC}{$v$} in \textcolor{noirC}{black}.}
	\label{fig:forcesComputation}
\end{figure}
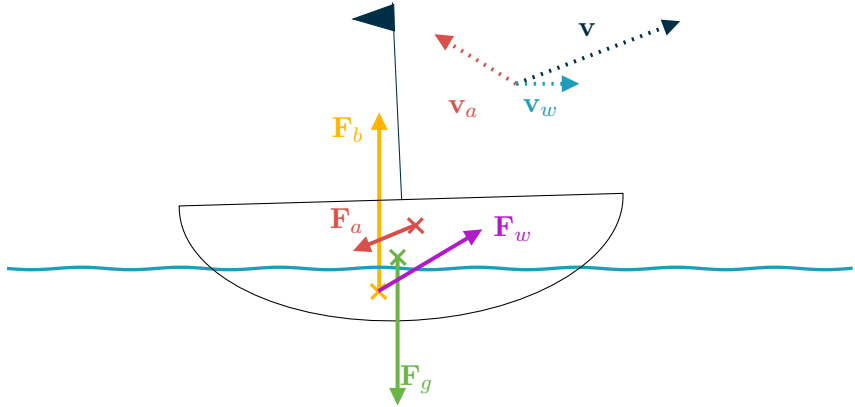

A summary of the forces applied to the solid is shown in Figure \ref{fig:forcesComputation}.

\section{Solid-to-fluid action}
\label{sec:solidToFluidAction}
\Framework relies on the method developed by Cords and Staadt (2009) \cite{cordsRealTimeOpenWater2009} to simulate the action of the solid on the fluid.
The simulation of this action is 3-dimensional by nature.
However, it has to be simplified to satisfy the third requirement of Section~\ref{sec:couplingSolidAndFluid}, concerning computation time performance.
As for fluid simulation in Section~\ref{sec:oceanWaves}, \framework assumes that the interaction from a solid to the fluid can be simulated as a 2D height field.
Furthermore, it considers a single frequency for the waves generated by a solid $M$.
These specific waves are called interactive waves.
These approximations may seem too strong, but as mentioned above, they are necessary to satisfy the requirements defined in Section~\ref{sec:couplingSolidAndFluid}.

Hence, \framework assumes that the interactive waves are solutions of the following 2D wave equation:
\begin{equation}
	\label{eq:2dwave}
	\Delta h(\mathbf x,t)-\frac{1}{c^2}\frac{\partial^2 h(\mathbf x,t)}{\partial t^2}=0
\end{equation}
where $h(\mathbf x,t)$ is the height of the water surface at horizontal position $\mathbf x$ and time $t$, $\Delta=\frac{\partial^2}{\partial x^2}+\frac{\partial^2}{\partial z^2}$ is the Laplacian in 2D and $c$ is the wave velocity expressed in $m.s^{-1}$.
\Framework uses the following Dirichlet boundary condition:
\begin{equation}
	h(\mathbf x,t) = 0 \text{~~~on the boundary.}
\end{equation}

\subsection{Finite-difference method}
\label{subsec:finiteDifferenceMethod}
Following Gomez (2000) \cite{gomezInteractiveSimulationWater2000},
\framework uses the FDM with an explicit scheme to numerically solve the equation (\ref{eq:2dwave}).
It uses a square simulation zone $Z$ of dimension $L\times L$ around a given mesh $M$.
This zone is regularly discretized with a $\delta$ step.
Time is also regularly discretized with a time step $dt$.

In the FDM simulation, \framework denotes by $h_{i,j}^n$ the water height $h(i\times \delta,j\times \delta,n \times dt)$ for discrete location $(i,j)\times \delta$ and time $n\times dt$.
This leads to the following explicit scheme:
\begin{equation}
	\label{eq:fdm_scheme}
	h_{i,j}^{n+1} = a\left(h_{i+1,j}^n+h_{i-1,j}^n+h_{i,j+1}^n+h_{i,j-1}^n-4h_{i,j}^n\right) + 2h_{i,j}^n - h_{i,j}^{n-1}
\end{equation}
where
\begin{equation}
	a=\frac{c^2dt^2}{\delta^2}.
\end{equation}

The Dirichlet boundary condition gives $h_{i,j}^n=0$ for $i$ or $j$ being on the boundary of $Z$. Since a new wave is generated at each time step, a damping effect is added to make them disappear gradually. Hence, the right part of Equation (\ref{eq:fdm_scheme}) is multiplied by the \textbf{damping factor} $d(t) = d^n$.

This damping factor is linearly related to the velocity of the mesh $\mathbf{v_M}(t)$ to ensure numerical stability and is defined as:
\begin{equation*}
	u = \text{clamp}\left(\frac{\|\mathbf{v_M}(t)\|}{v_{max}}, 0, 1\right),
\end{equation*}
\begin{equation}
	d(t) = \text{lerp}(d_0, d_{max}, u),
\end{equation}
where:
\begin{itemize}
	\item $\text{clamp}(x, 0, 1)$ limits the value of $x$ to the range $[0, 1]$;
	\item $\text{lerp}(a, b, t) = (1-t)a + tb$ is the linear interpolation function;
	\item $v_{max}$ is the velocity threshold, above which the damping factor equals its maximum value $d_{max}$;
	\item $d_0$ is the damping value for a null velocity.
\end{itemize}

For a plausible appearance, \framework uses the following values by default:
\begin{equation}
	\begin{cases}
		d_0     & = 0.98,       \\
		d_{max} & = 0.999,      \\
		v_{max} & = 5 m.s^{-1}.
	\end{cases}
\end{equation}
Nevertheless, the user can adjust these values in a way that a body $M$ causes more or less interaction waves.

\subsection{Grid translation}
\label{subsec:gridTranslation}
The waves generated by a given body $M$ are defined in its local coordinates system.
Hence, when $M$ moves, the waves have to move accordingly as done by Cords and Staadt (2009) \cite{cordsRealTimeOpenWater2009}.
A first solution consists of applying the translation to the grid at times $t$ and $t-dt$ and then of using the integration scheme of Equation (\ref{eq:fdm_scheme}).
Nevertheless, these time-consuming shifts can be avoided by using the following domain translation:
\begin{equation}
	\begin{cases}
		k = & i - \left\lfloor\frac{\mathbf{p}_x(t+dt)-\mathbf{p}_x(t)}{\delta}\right\rfloor    \\
		l = & j - \left\lfloor\frac{\mathbf{p}_z(t+dt)-\mathbf{p}_z(t)}{\delta}\right\rfloor    \\
		o = & i - \left\lfloor\frac{\mathbf{p}_x(t+dt)-\mathbf{p}_x(t-dt)}{\delta}\right\rfloor \\
		p = & j - \left\lfloor\frac{\mathbf{p}_z(t+dt)-\mathbf{p}_z(t-dt)}{\delta}\right\rfloor
	\end{cases}
\end{equation}

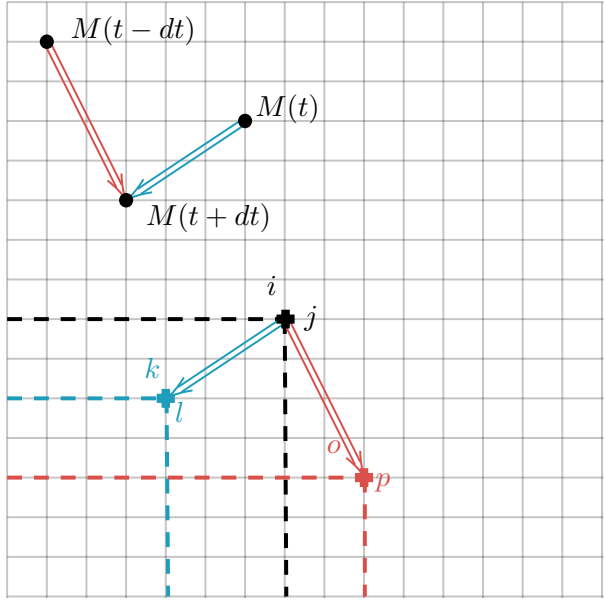
\begin{figure}[!h]
	\centering
	\tikzset{every picture/.style={line width=0.75pt}} 

\begin{tikzpicture}[x=0.75pt,y=0.75pt,yscale=-1,xscale=1]

\draw  [draw opacity=0][fill={rgb, 255:red, 0; green, 0; blue, 0 }  ,fill opacity=0 ][line width=0.75]  (129,35.4) -- (429.8,35.4) -- (429.8,336.2) -- (129,336.2) -- cycle ; \draw  [color={rgb, 255:red, 0; green, 0; blue, 0 }  ,draw opacity=0.23 ][line width=0.75]  (129,35.4) -- (129,336.2)(149,35.4) -- (149,336.2)(169,35.4) -- (169,336.2)(189,35.4) -- (189,336.2)(209,35.4) -- (209,336.2)(229,35.4) -- (229,336.2)(249,35.4) -- (249,336.2)(269,35.4) -- (269,336.2)(289,35.4) -- (289,336.2)(309,35.4) -- (309,336.2)(329,35.4) -- (329,336.2)(349,35.4) -- (349,336.2)(369,35.4) -- (369,336.2)(389,35.4) -- (389,336.2)(409,35.4) -- (409,336.2)(429,35.4) -- (429,336.2) ; \draw  [color={rgb, 255:red, 0; green, 0; blue, 0 }  ,draw opacity=0.23 ][line width=0.75]  (129,35.4) -- (429.8,35.4)(129,55.4) -- (429.8,55.4)(129,75.4) -- (429.8,75.4)(129,95.4) -- (429.8,95.4)(129,115.4) -- (429.8,115.4)(129,135.4) -- (429.8,135.4)(129,155.4) -- (429.8,155.4)(129,175.4) -- (429.8,175.4)(129,195.4) -- (429.8,195.4)(129,215.4) -- (429.8,215.4)(129,235.4) -- (429.8,235.4)(129,255.4) -- (429.8,255.4)(129,275.4) -- (429.8,275.4)(129,295.4) -- (429.8,295.4)(129,315.4) -- (429.8,315.4)(129,335.4) -- (429.8,335.4) ; \draw  [color={rgb, 255:red, 0; green, 0; blue, 0 }  ,draw opacity=0.23 ][line width=0.75]   ;
\draw [color=bleuF  ,draw opacity=1 ]   (194.82,129.71) -- (248.17,94.15)(196.49,132.21) -- (249.83,96.65) ;
\draw [shift={(189,135.4)}, rotate = 326.31] [color=bleuF  ,draw opacity=1 ][line width=0.75]    (13.12,-3.95) .. controls (8.34,-1.68) and (3.97,-0.36) .. (0,0) .. controls (3.97,0.36) and (8.34,1.68) .. (13.12,3.95)   ;

\draw [color=redF  ,draw opacity=1 ]   (184.08,128.92) -- (147.66,56.07)(186.76,127.57) -- (150.34,54.73) ;
\draw [shift={(189,135.4)}, rotate = 243.43] [color=redF  ,draw opacity=1 ][line width=0.75]    (13.12,-3.95) .. controls (8.34,-1.68) and (3.97,-0.36) .. (0,0) .. controls (3.97,0.36) and (8.34,1.68) .. (13.12,3.95)   ;
\draw [line width=1.5]  [dash pattern={on 5.63pt off 4.5pt}]  (269,195.4) -- (269.8,335.4) ;
\draw [line width=1.5]  [dash pattern={on 5.63pt off 4.5pt}]  (129,195.4) -- (269.4,195.4) ;
\draw [color=bleuF  ,draw opacity=1 ]   (215.22,229.71) -- (268.57,194.15)(216.89,232.21) -- (270.23,196.65) ;
\draw [shift={(209.4,235.4)}, rotate = 326.31] [color=bleuF  ,draw opacity=1 ][line width=0.75]    (13.12,-3.95) .. controls (8.34,-1.68) and (3.97,-0.36) .. (0,0) .. controls (3.97,0.36) and (8.34,1.68) .. (13.12,3.95)   ;
\draw [color=bleuF  ,draw opacity=1 ][line width=1.5]  [dash pattern={on 5.63pt off 4.5pt}]  (209.4,235.4) -- (210.2,335.4) ;
\draw [color=bleuF  ,draw opacity=1 ][line width=1.5]  [dash pattern={on 5.63pt off 4.5pt}]  (129,235.4) -- (209.4,235.4) ;
\draw [color=redF  ,draw opacity=1 ]   (304.48,268.92) -- (268.06,196.07)(307.16,267.57) -- (270.74,194.73) ;
\draw [shift={(309.4,275.4)}, rotate = 243.43] [color=redF  ,draw opacity=1 ][line width=0.75]    (13.12,-3.95) .. controls (8.34,-1.68) and (3.97,-0.36) .. (0,0) .. controls (3.97,0.36) and (8.34,1.68) .. (13.12,3.95)   ;
\draw [color=redF  ,draw opacity=1 ][line width=1.5]  [dash pattern={on 5.63pt off 4.5pt}]  (309.4,275.4) -- (309.4,335.4) ;
\draw [color=redF  ,draw opacity=1 ][line width=1.5]  [dash pattern={on 5.63pt off 4.5pt}]  (129,275.4) -- (310.2,275.4) ;
\draw  [draw opacity=0][fill={rgb, 255:red, 0; green, 0; blue, 0 }  ,fill opacity=1 ] (185.45,135.4) .. controls (185.45,133.38) and (187.04,131.75) .. (189,131.75) .. controls (190.96,131.75) and (192.55,133.38) .. (192.55,135.4) .. controls (192.55,137.42) and (190.96,139.05) .. (189,139.05) .. controls (187.04,139.05) and (185.45,137.42) .. (185.45,135.4) -- cycle ;
\draw  [draw opacity=0][fill={rgb, 255:red, 0; green, 0; blue, 0 }  ,fill opacity=1 ] (145.45,55.4) .. controls (145.45,53.38) and (147.04,51.75) .. (149,51.75) .. controls (150.96,51.75) and (152.55,53.38) .. (152.55,55.4) .. controls (152.55,57.42) and (150.96,59.05) .. (149,59.05) .. controls (147.04,59.05) and (145.45,57.42) .. (145.45,55.4) -- cycle ;
\draw  [draw opacity=0][fill={rgb, 255:red, 0; green, 0; blue, 0 }  ,fill opacity=1 ] (245.45,95.4) .. controls (245.45,93.38) and (247.04,91.75) .. (249,91.75) .. controls (250.96,91.75) and (252.55,93.38) .. (252.55,95.4) .. controls (252.55,97.42) and (250.96,99.05) .. (249,99.05) .. controls (247.04,99.05) and (245.45,97.42) .. (245.45,95.4) -- cycle ;
\draw  [fill={rgb, 255:red, 0; green, 0; blue, 0 }  ,fill opacity=1 ][line width=0.75]  (267.77,191.15) -- (271.03,191.15) -- (271.03,193.59) -- (273.48,193.59) -- (273.48,197.21) -- (271.03,197.21) -- (271.03,199.65) -- (267.77,199.65) -- (267.77,197.21) -- (265.33,197.21) -- (265.33,193.59) -- (267.77,193.59) -- cycle ;
\draw  [color=bleuF  ,draw opacity=1 ][fill=bleuF  ,fill opacity=1 ][line width=0.75]  (207.37,231.15) -- (210.63,231.15) -- (210.63,233.59) -- (213.07,233.59) -- (213.07,237.2) -- (210.63,237.2) -- (210.63,239.65) -- (207.37,239.65) -- (207.37,237.2) -- (204.92,237.2) -- (204.92,233.59) -- (207.37,233.59) -- cycle ;
\draw  [color=redF  ,draw opacity=1 ][fill=redF  ,fill opacity=1 ][line width=0.75]  (307.37,271.15) -- (310.63,271.15) -- (310.63,273.59) -- (313.07,273.59) -- (313.07,277.2) -- (310.63,277.2) -- (310.63,279.65) -- (307.37,279.65) -- (307.37,277.2) -- (304.92,277.2) -- (304.92,273.59) -- (307.37,273.59) -- cycle ;

\draw (197.6,135.4) node [anchor=north west][inner sep=0.75pt]   [align=left] {$\displaystyle M( t+dt)$};
\draw (252.8,80.2) node [anchor=north west][inner sep=0.75pt]   [align=left] {$\displaystyle M( t)$};
\draw (159.6,41) node [anchor=north west][inner sep=0.75pt]   [align=left] {$\displaystyle M( t-dt)$};
\draw (258.4,172.2) node [anchor=north west][inner sep=0.75pt]   [align=left] {$\displaystyle i$};
\draw (277.47,187.4) node [anchor=north west][inner sep=0.75pt]   [align=left] {$\displaystyle j$};
\draw (212.28,235.8) node [anchor=north west][inner sep=0.75pt]  [color=bleuF  ,opacity=1 ] [align=left] {$\displaystyle l$};
\draw (196.8,213.4) node [anchor=north west][inner sep=0.75pt]  [color=bleuF  ,opacity=1 ] [align=left] {$\displaystyle k$};
\draw (288.68,255) node [anchor=north west][inner sep=0.75pt]  [color=redF  ,opacity=1 ] [align=left] {$\displaystyle o$};
\draw (313.48,271.8) node [anchor=north west][inner sep=0.75pt]  [color=redF  ,opacity=1 ] [align=left] {$\displaystyle p$};

\end{tikzpicture}
	\caption{An example of the relationship between the indices $(i,j)$, $(k,l)$, and $(o,p)$ in the finite-difference method grid translation, showing how the grid indices are affected by the motion of the body $M$ at different time steps $t$, $t-dt$, and $t+dt$}
	\label{fig:mdfIndices}
\end{figure}

where $\mathbf p(t)=\left[p_x(t)\ p_z(t)\right]^T$ is the position of the mesh $M$ at time $t$, $k$ and $l$ (respectively $o$ and $p$) are the shifted indices (see Figure \ref{fig:mdfIndices}) defined by the translation of $M$ at times $t+dt$ from times $t$ (respectively from times $t-dt$).
Then, the integration scheme can be rewritten as follows:
\begin{equation}
	\label{eq:fdm_scheme_corrected}
	h_{i,j}^{n+1} = d^n\left(a\left(h_{k+1,l}^n+h_{k-1,l}^n+h_{k,l+1}^n+h_{k,l-1}^n- 4 h_{k,l}^n\right) + 2 h_{k,l}^n - h_{o,p}^{n-1}\right)
\end{equation}

\begin{figure}[!h]
	\centering
	\input{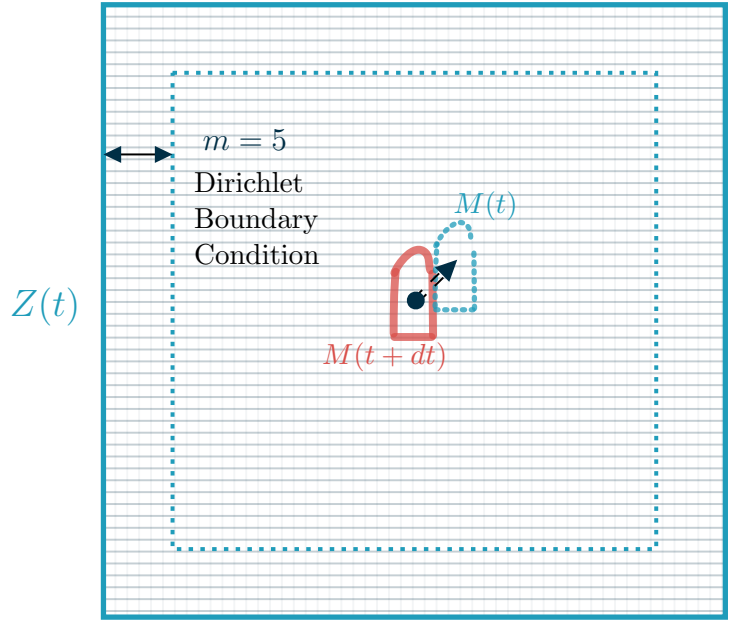}
	\caption{An example of zone $Z(t)$ at a time $t$ using boundary size $m=5$.
		In \textcolor{bleuF}{dotted blue} the mesh ${M}$ at time $t$.
		In \textcolor{redF}{red} the same mesh at time $t+dt$.}
	\label{fig:examGrid}
\end{figure}

Since $M$ moves, the shift indexes $k$, $l$, $o,$ and $p$ can be outside the boundary of the simulation zone $Z$.
To solve this problem, the size $m$ of the boundary of $Z$ is increased such that $m>1$ (see Figure \ref{fig:examGrid}).
\Framework takes $m=16$ which is large enough for large translations.

To ensure high performance, the FDM update step is executed entirely on the GPU, with a single kernel processing all vertices of the FDM grid in parallel.
Each thread updates one grid point, significantly reducing computation time. The pseudocode associated to this method is shown in Algorithm \ref{alg:fdm_compute}.

\begin{algorithm}
\caption{FDM Algorithm}
\label{alg:fdm_compute}
\begin{algorithmic}[1]
\Require List of zone-specific parameters: $a$, $b$, $d(t)$, $\mathbf{p}(t-dt)$, $\mathbf{p}(t)$ and $\mathbf{p}(t-dt)$.
\For{each vertices of the FDM grid $i,j$ and zone $Z$ in the simulation domain}
    \State $k = i - \left\lfloor\frac{\mathbf{p}_x(t+dt)-\mathbf{p}_x(t)}{\delta}\right\rfloor$
    \State $l = j - \left\lfloor\frac{\mathbf{p}_z(t+dt)-\mathbf{p}_z(t)}{\delta}\right\rfloor $
    \State $o = i - \left\lfloor\frac{\mathbf{p}_x(t+dt)-\mathbf{p}_x(t-dt)}{\delta}\right\rfloor$
    \State $p = j - \left\lfloor\frac{\mathbf{p}_z(t+dt)-\mathbf{p}_z(t-dt)}{\delta}\right\rfloor$
    \State $h_{i,j}^{n+1} = d \cdot \big(a \cdot (h_{k-1,l} + h_{k+1,l} + h_{k,l-1} + h_{k,l+1})+ b \cdot h_{k,l} - h_{o,p}^{l-1} \big)$
\EndFor
\end{algorithmic}
\end{algorithm}

\subsection{Stability}
\label{subsec:stability}
The stability of the scheme defined by Equation (\ref{eq:fdm_scheme_corrected}) is related to the CFL condition (for Courant–Friedrichs–Lewy), here given as follows:
\begin{equation}
	\label{equa:cfl}
	\frac{c^2dt^2}{\delta^2}\leq 0.5
\end{equation}
The parameter $dt$ is considered fixed throughout the simulation.
One contribution of this article is to adapt the simulation's parameters $c$ and $\delta$ to the CFL, as presented below.

Preliminary tests have shown that $\delta$ must be small for a low velocity $\mathbf{v_M}(t)$ of $M$.
Moreover, $\delta$ must be large enough for $Z$ to encompass $M$ and the waves generated by $M$.
The size of $M$ is defined from its bounding box $(b_x, b_y, b_z)$ as $S(M)=\max(b_x, b_z)$.
\Framework lets the user define the bounds $\delta_{min}$ and $\delta_{max}$ so that $Z$ is at least twice as large as $M$.
Then $\delta$ is defined as follows:
\begin{equation}
	\delta =
	\begin{cases}
		0.999~dt                    & ~~ \text{if}~||\mathbf{v_M}(t)||<1    \\
		||\mathbf{v_M}(t)||0.999~dt & ~~ \text{if}~||\mathbf{v_M}(t)||\geq1
	\end{cases}
\end{equation}
Finally $\delta$ is clamped between $\delta_{min}$ and $\delta_{max}$.
The value of $c$ is deduced from $\delta$ and $dt$ to ensure that the inequality in Equation \ref{equa:cfl} is always satisfied:
\begin{equation}
	c = \sqrt{0.49} \frac{\delta}{dt}.
\end{equation}
This choice of $c$ ensures numerical stability.

\subsection{Wave generation}
\label{subsec:mask}
While the FDM allows to simulate the propagation of the interactive waves, one more input is required for the wave generation: an alteration of the free surface of the height map.
\Framework uses a new way to compute this input, which provides a satisfying result no matter the form of the body generating the waves.
The input of the FDM is called \textbf{mask}.

The idea behind the mask is to distort the grid of the FDM by reproducing the intersection surface between $M$ and the free surface.
Here, free surface refers to the ambient waves of Tessendorf's method (see Section \ref{sec:oceanWaves}) plus the interactive waves generated by all solids except $M$.
It does \textbf{not} include the free surface generated by $M$, or the simulation would face a divergence issue.

This intersection has already been computed as part of the geometric parameters in Section~\ref{subsec:geomParam}.
It is given as a list of 3D points forming the intersection polygons of the hull with the free surface.
These points are moved from the world coordinates system to the one of $Z$, the surrounding grid of $M$.

Hence, for a body $M$ the wave generation relies on two different steps:
First, a mask is defined from this list of 3D points, corresponding to the vertices of $Z$ contained in one intersected polygons.
Second, the height $h$ of each point in the mask is determined using an ad hoc function.
These two steps are detailed in the following two sections.

\subsubsection{Calculate the mask}
The FDM needs the height of the vertices below the body.
This is computed from the mask, composed of the vertices of $Z$ contained in one of the intersections polygons at least.
This is done using the \textbf{Point in Polygon} strategy described in Haines (1994) \cite{hainesPointPolygonStrategies1994}.

The principle of this algorithm is as follows:
for each vertex $v$ of the grid $Z$, a ray starting at $v$ and with direction $z$ is generated, and its number of intersection with all the intersection polygons is computed.
$v$ is in the polygon if this number of intersections is odd; otherwise it is not.

To ensure consistency regardless of the body's orientation, a rotation alignment is applied to the intersected polygons before processing.

To reduce the computation times, this strategy is applied to the vertices $v$ included into the bounding box of the intersections polygons.
Since the number of vertices $v$ of $Z$ is at least twice as the size of $M$, this simple method reduces efficiently the number of ray to polygon calculations.

\subsubsection{Evaluate the height of vertices in the mask}
\label{subsubsec:heightMask}
\begin{figure}[htbp]
	\centering

	\subfloat[Three quarter view]
	{
		\includegraphics[width=0.25\textwidth]{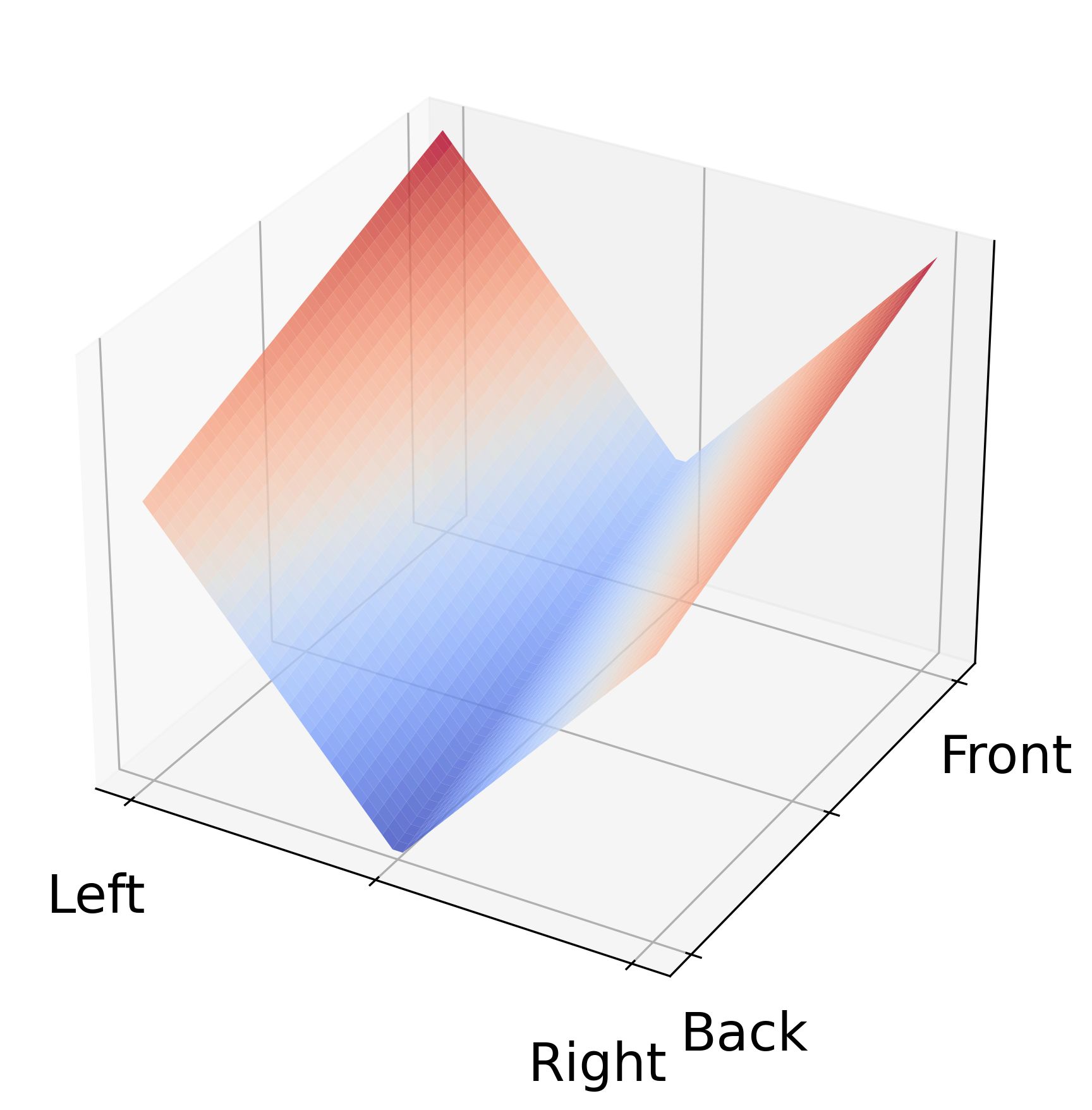}
	}
	\hspace{1cm}
	\subfloat[Back view]
	{
		\includegraphics[width=0.25\textwidth]{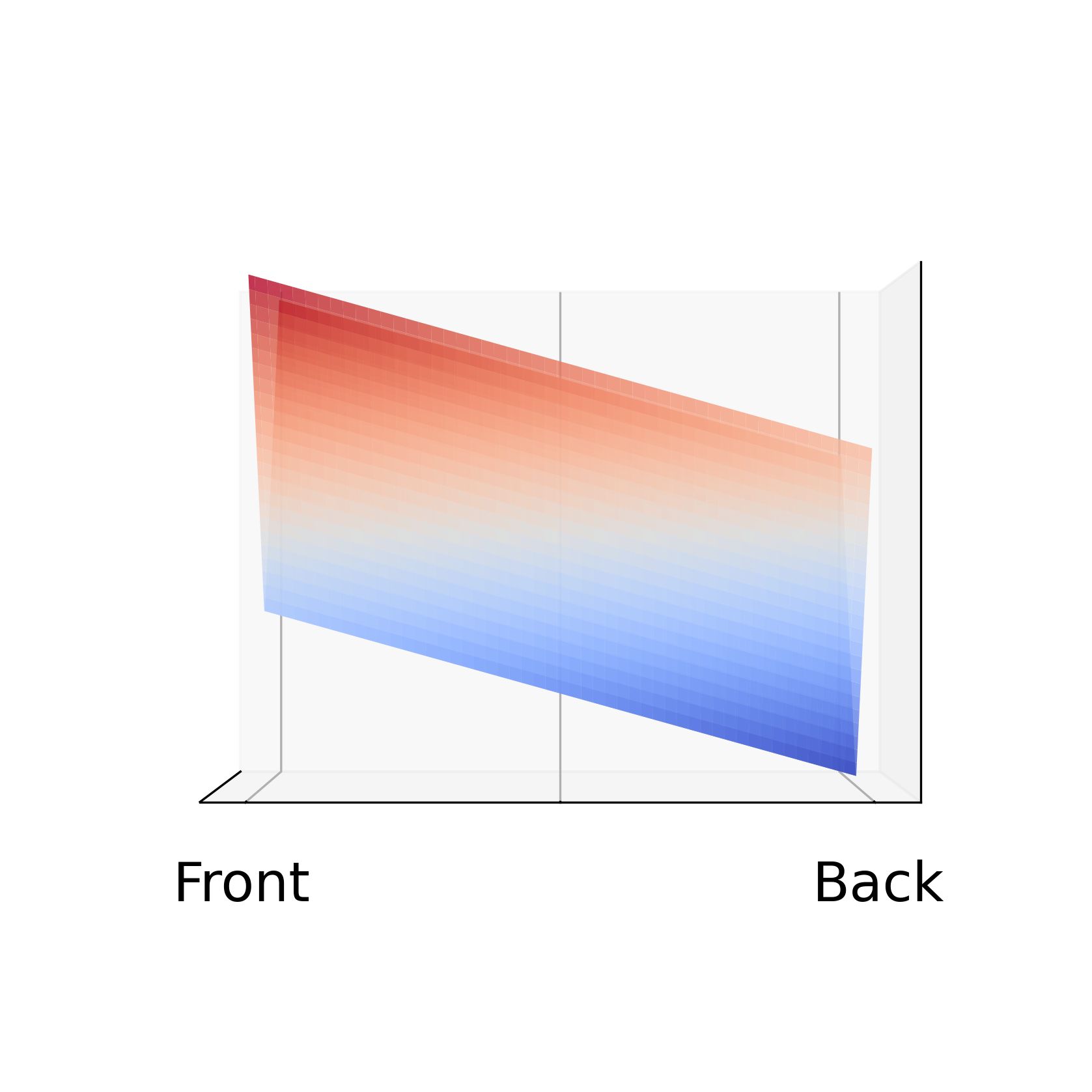}
	}
	\hspace{1cm}
	\subfloat[Side view]
	{
		\includegraphics[width=0.25\textwidth]{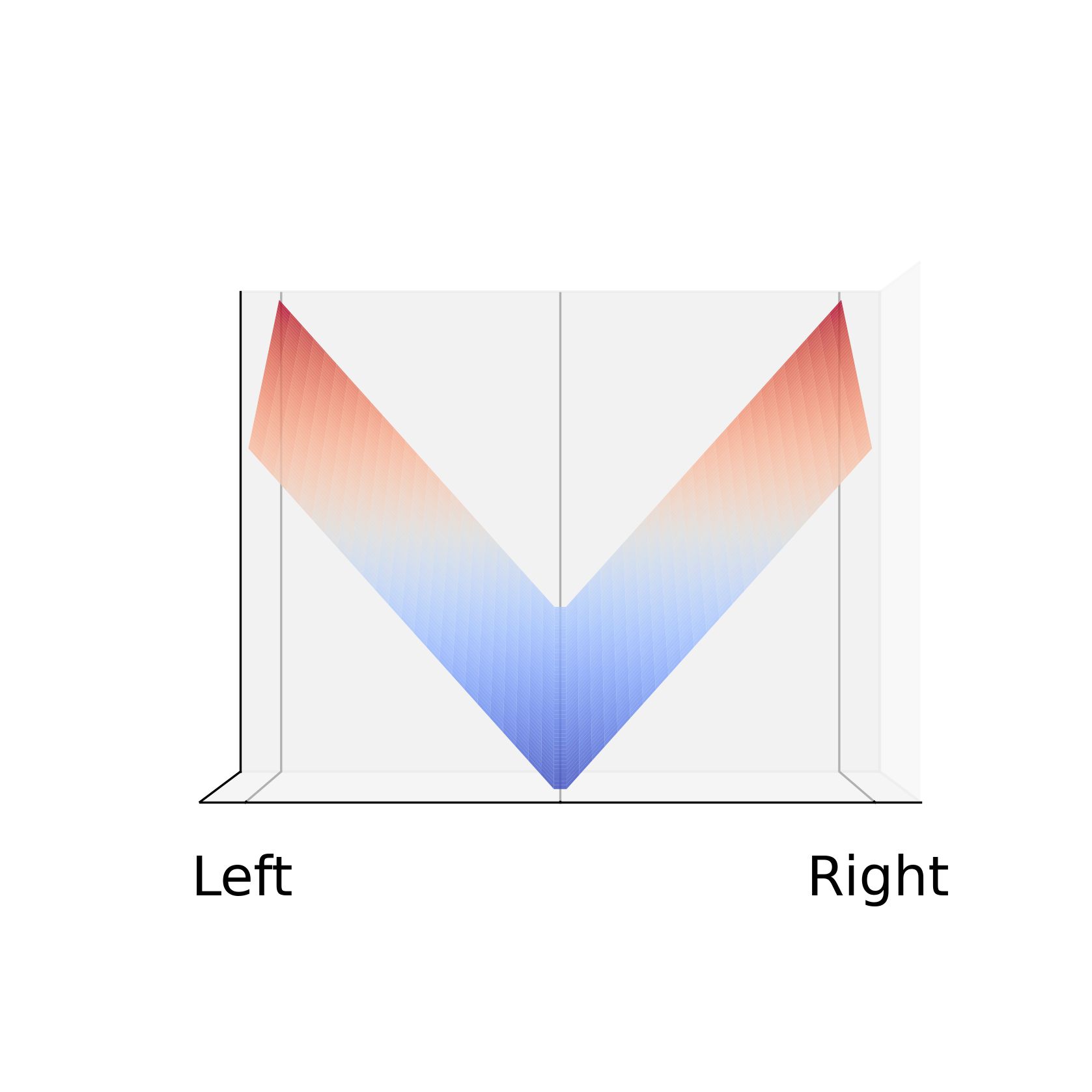}
	}
	\caption{Three different theoretical point of view of the mask for a simple rectangle.}
	\label{fig:mask_maskTheoreticalView}
\end{figure}

Knowing all the vertices in the mask, their height must be given as input of the FDM.
\Framework uses a simple model that tries to reproduce parts of the idea of Cords and Staadt (2009) \cite{cordsRealTimeOpenWater2009}.
By construction of the grid $Z$, $M$ has its bow oriented on the positive side of the $z$ axis, and its starboard (right-hand side when facing the bow) on the positive side of the $x$ axis.
Hence, $h$ is made such that from the vertex at position $\mathbf x$, the front of the mask is above the free surface and the back below, in a way that is proportional to the speed of $M$.
Moreover, the boat forms a $V$ shape laterally, with its center a bit below the extremity of the side. Figure \ref{fig:mask_maskTheoreticalView} presents the desired shape from three different perspectives.

To simplify, $h(\mathbf x)$ is defined as the sum of two orthogonal functions $f(x)$ and $g(z)$,
where $f(x)$ represent the side effect, and $g(z)$ the front/back effect.

The function $f$ models the side effect using a shape in $V$.
It is defined as follows:
\begin{equation}
	f(x) = \frac{\left|x-c_x\right|}{b_x}.
\end{equation}
The $V$-shape is obtained from the absolute value of the abscissa $x$ shifted to the center $c_x$ of $M$.
The amplitude of the height is controlled by the inverse of the bounding box size $b_x$ of $M$ on the $x$ axis.

The purpose of $g$ is to shift up the front of the interactive wave and shift down its back.
The idea is that a moving body displaces the water from its submerged hull, producing a high interactive wave in front of the body.
This effect is modeled using a quite simple affine function $g(z)=a\cdot z+b$.
The slope coefficient $a$ is defined considering the maximum height $h_b$ on the back of the mask and the height $h_f$ on its front.

While $h_b$ is defined by the user, the value of $h_f$ is estimated as follows:
\begin{equation}
	h_f = ||\mathbf{v_M}||h_{M}\times i_f\times\frac{v_w}{v_{M}},
\end{equation}
where $\mathbf{v_M}$ is the velocity of the body $M$, $h_M$ is the full height of $M$, $i_f$ is a user parameter allowing more user control, $v_w$ is the submerged volume of $M$ (see Section~\ref{sec:fluidToSolidAction}), and $v_{M}$ is the full volume of $M$.
Hence, for more realism $h_f$ is made proportional to the velocity of $M$:
the faster $M$ moves, the bigger the waves generated by $M$ at its front are.

The values of $a$ and $b$ are straightforward to obtain by considering the height at the back and the front, so using the relations $g(z_{min})=h_b$ and $g(z_{max})=h_f$,
where $z_{min}$ is the $z$-coordinates of the back and $z_{max}$ the $z$-coordinate of the front.
From the first relation, it comes $b=h_b-a\cdot z_{min}$.
From the second, it comes $a=\left(h_f-h_b\right)/b_z$, where $b_z=z_{max}-z_{min}$ is the bounding box extend in $z$ axis.
This allows to simplify the term $b$ as $b=\left(h_bz_{max}-h_fz_{min}\right)/b_z$.

The functions $f(x)$ and $g(z)$ are then used to obtain the height of the mask as follows,
but using the user factor $b_w$ to allow more user control:
\begin{equation}
	\label{equa:mask_full}
	h(\mathbf x)=b_w\left(f(x)+g(z)\right).
\end{equation}

Note that $h$ is only well-defined if the bow of $M$ is oriented on the positive $z$ axis.
Therefore, a 2D rotation is first applied to the vertices $\mathbf x$ before applying $h$.

\begin{figure}[htbp]
	\centering
	\subfloat[The height of the mask of a simple boat encoded in a single channel texture.]
	{
		\includegraphics[width=0.4\textwidth]{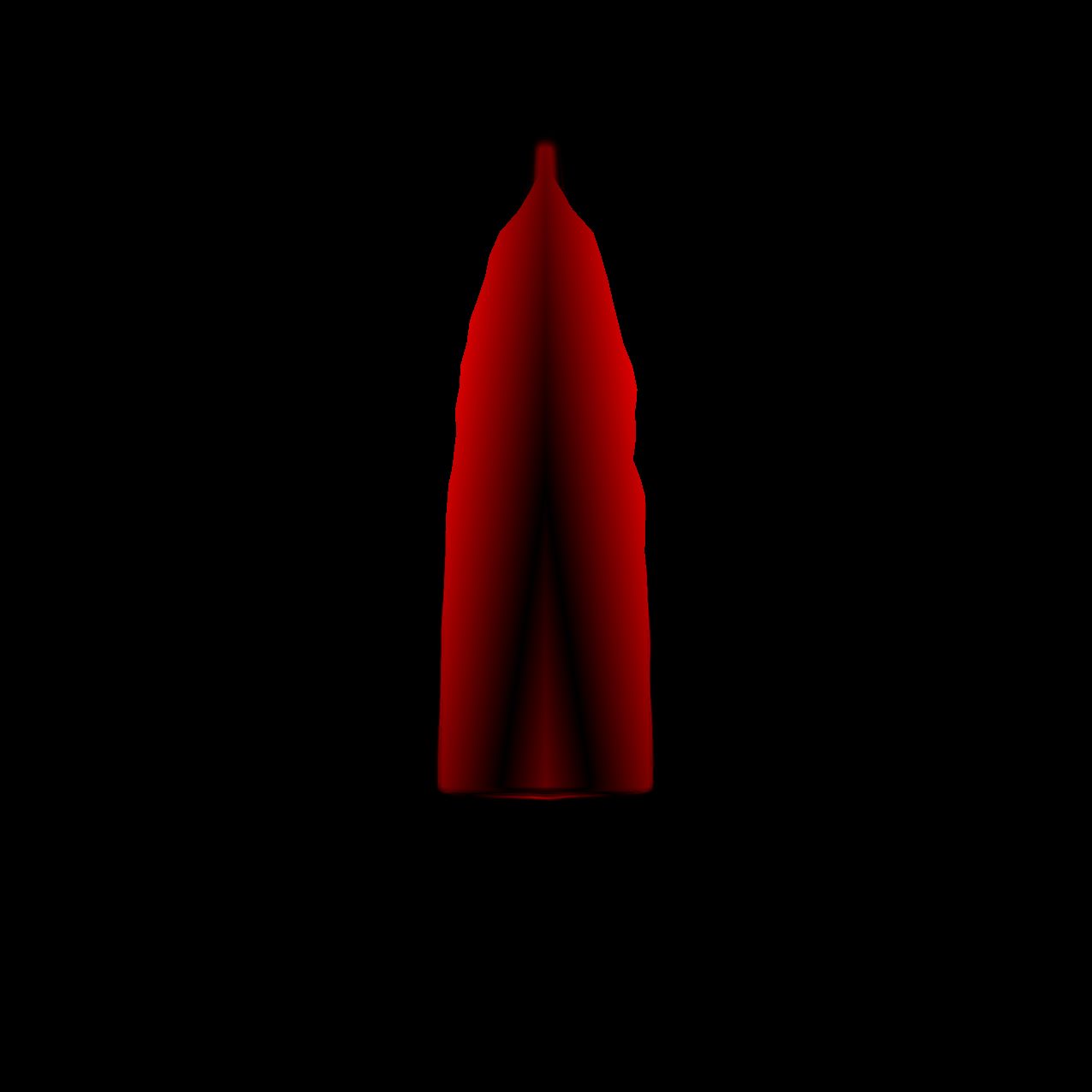}
		\label{subfig:maskHeightMap}
	}
	\hspace{1cm}
	\subfloat[The height of the mask of a simple boat in a 3D scene with a very low sea state. Boat rendering and wave propagation has been disabled to only display the initial step of the height of the mask.]
	{
		\includegraphics[width=0.4\textwidth]{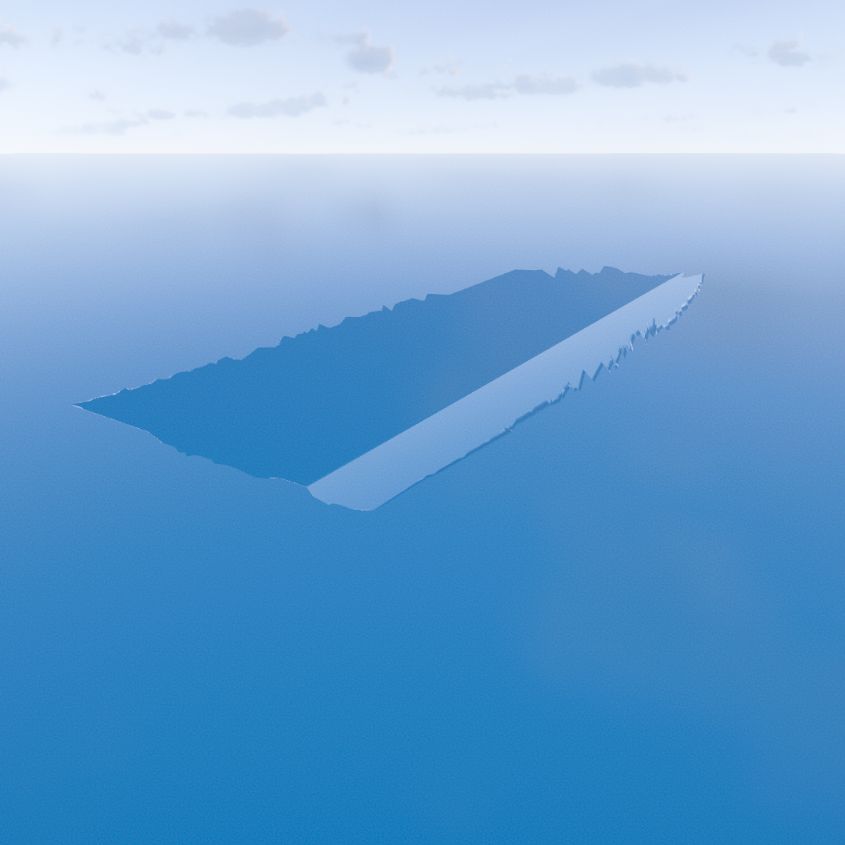}
		\label{subfig:mask3D}
	}
	\caption{Two different views of the height of the mask defined by Equation \ref{equa:mask_full} for a simple boat.}
	\label{fig:mask}
\end{figure}

Figure \ref{fig:mask} shows some views of the application of this interactive waves process.

The mask calculation is implemented on the GPU to leverage parallel processing efficiency. A dedicated kernel is launched for each zone only for the vertices of the FDM grid that fall within the bounding box of the intersection polygon, reducing computational overhead. Each thread independently performs the point-in-polygon test using a ray-intersection method and determines whether the corresponding vertex should be included in the mask. Once identified as inside an intersection polygon, the height function is applied per vertex, using the predefined wave generation model. The full process is detailed in Algorithm \ref{alg:mask_calc}.

\begin{algorithm}
\caption{Mask Calculation via Point-in-Polygon Test}
\label{alg:mask_calc}
\begin{algorithmic}[1]
\Require The bounding box, the intersection polygon
\For{each vertices of the FDM grid $i,j$ within the bounding box in a zone $Z$} 
    \State  $counter \gets 0$ \Comment{Initialize the counter for intersections}
    \For{each edges $(v_k,v_{k+1})$ of the intersection polygon}
        \State $p_1 = (i,j)$ 
        \State $p_2 = (i+\infty,j+\infty)$ 
        \State $p_3 = v_k$
        \State $p_4 = v_{k+1}$
        \If{Segment-Intersect($p_1$, $p_2$, $p_3$, $p_4$) do}
            \State $counter+=1$
        \EndIf
    \EndFor
    \If{$counter \bmod 2 = 1$} \Comment{Point $(i,j)$ is inside an intersection polygon.}
        \State $h_{i,j}^{n} = b_w\left(f(x)+g(z)\right)$ \Comment{See Equation \ref{equa:mask_full}}
    \EndIf
\EndFor     
\end{algorithmic}
\end{algorithm}

\section{Results and discussions}
\label{sec:results_and_discussions}

\begin{figure}
	\centering
	\tikzset{every picture/.style={line width=0.75pt}} 
	\input{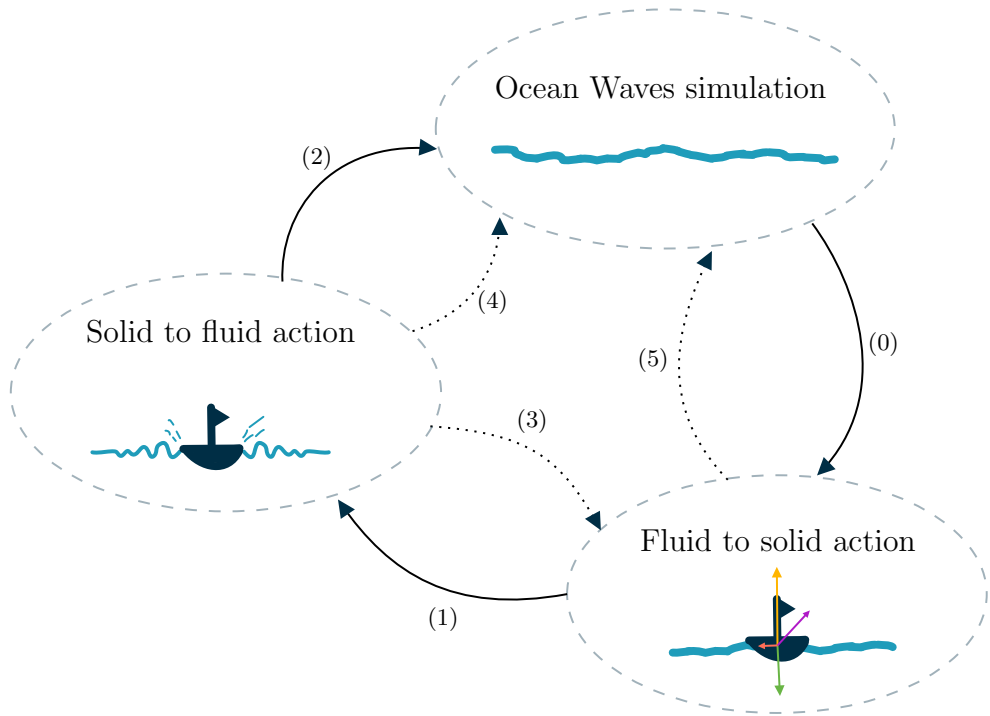}
	\caption{Interactions between the ocean waves simulation (Section~\ref{sec:oceanWaves}), the fluid-to-solid method (Section~\ref{sec:fluidToSolidAction}) and the solid-to-fluid method (Section~\ref{sec:solidToFluidAction}).
		The arrow represents an interaction between two methods.
		The arrow direction indicates that the method from which the arrow originates impacts the method to which the arrow is pointing to.}
	\label{fig:interactionMethods}
\end{figure}

Sections \ref{sec:oceanWaves}, \ref{sec:fluidToSolidAction} and \ref{sec:solidToFluidAction} describe three methods to handle the main parts of an ocean simulation: the free surface, and the coupling between a solid and a fluid.
Each method has been chosen to find the best ratio between high performance and physics accuracy.

Notice that each method interacts naturally with the two others, as shown in Figure~\ref{fig:interactionMethods}.
Indeed, Tessendorf's method generates the free surface.
As depicted by arrow $(0)$, the free surface method impacts the force calculation of the fluid-to-solid method.
The force calculation of the fluid-to-solid method impacts the mask calculation (arrow $(1)$), which is the input of the solid-to-fluid method.
The solid-to-fluid method modifies the free surface around the solid and is an input of the ocean waves simulation (arrow $(2)$).
All these interactions loops through the simulation over time.
As a result, the three components of \framework have an impact on the two others (arrows $(3)$, $(4)$ and $(5)$) at the next time-step.
For example, since the solid-to-fluid method impacts the free surface (arrow $(2)$), and the free surface impacts the force calculation used in the fluid-to-solid (arrow $(0)$), then indirectly the solid-to-fluid method has an indirect impact on the fluid-to-solid action (arrow $(3)$).

Two experiments with different solids have been conducted to demonstrate the efficiency of \framework.
These experiments ran on a \href{https://www.intel.fr/content/www/fr/fr/products/sku/199316/intel-core-i710700-processor-16m-cache-up-to-4-80-ghz/specifications.html}{Intel Core i7-10700} and a \href{https://www.nvidia.com/fr-fr/geforce/graphics-cards/compare/?section=compare-20}{Nvidia GeForce RTX 2070 Super} using single-precision floating-point numbers.
The ocean was simulated with three cascades using a resolution of $256\times265$.
Moreover, the resolution of FDM grids is set to $512\times 512$.
\begin{figure}[htbp]
	\centering

	\subfloat[\emph{One-solid} using sea state $1$.]
	{
		\includegraphics[width=0.4\textwidth]{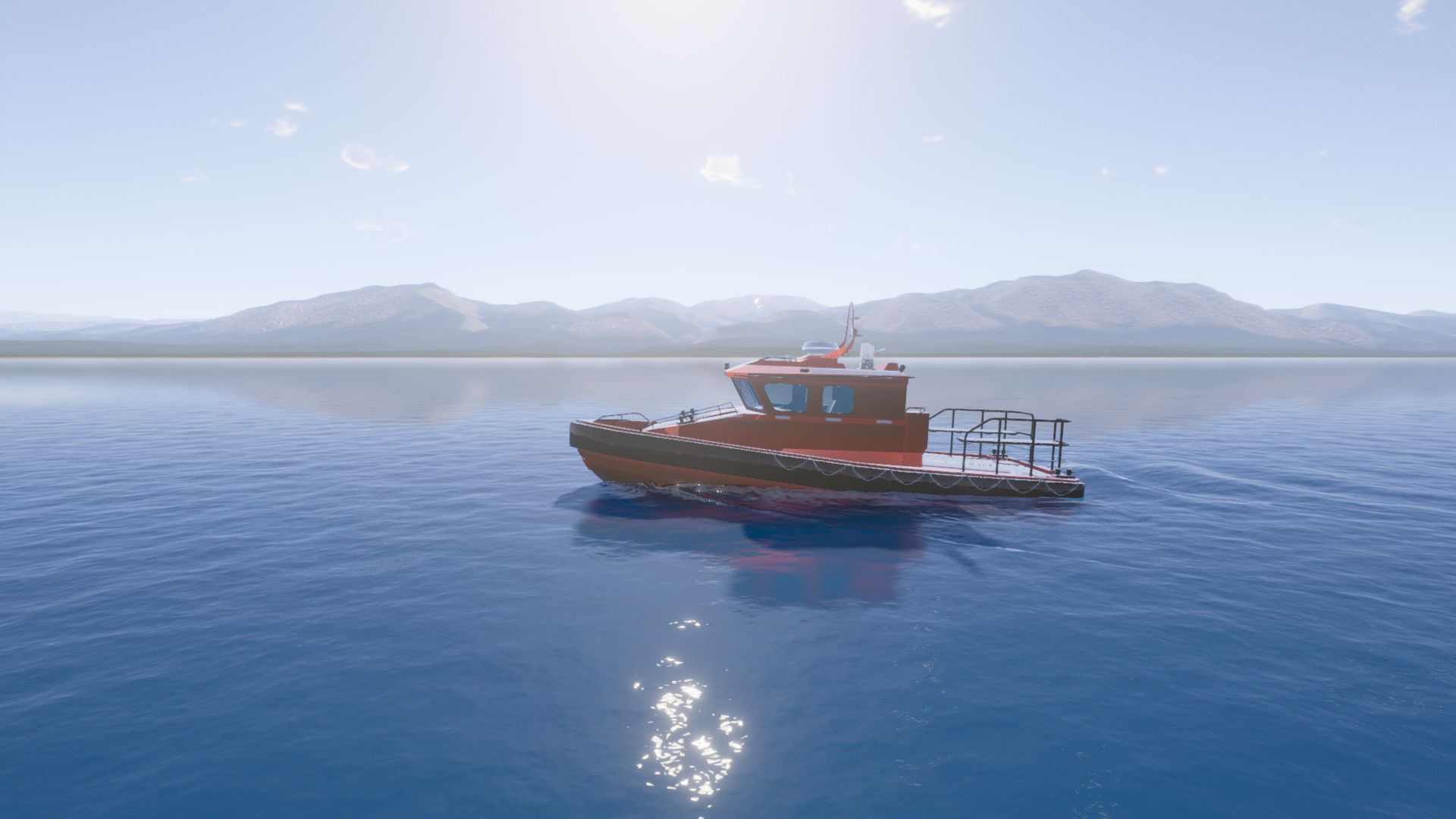}
		\label{fig:experiment1-1}
	}
	\hspace{1cm}
	\subfloat[\emph{One-solid} using sea state $6$.]
	{
		\includegraphics[width=0.4\textwidth]{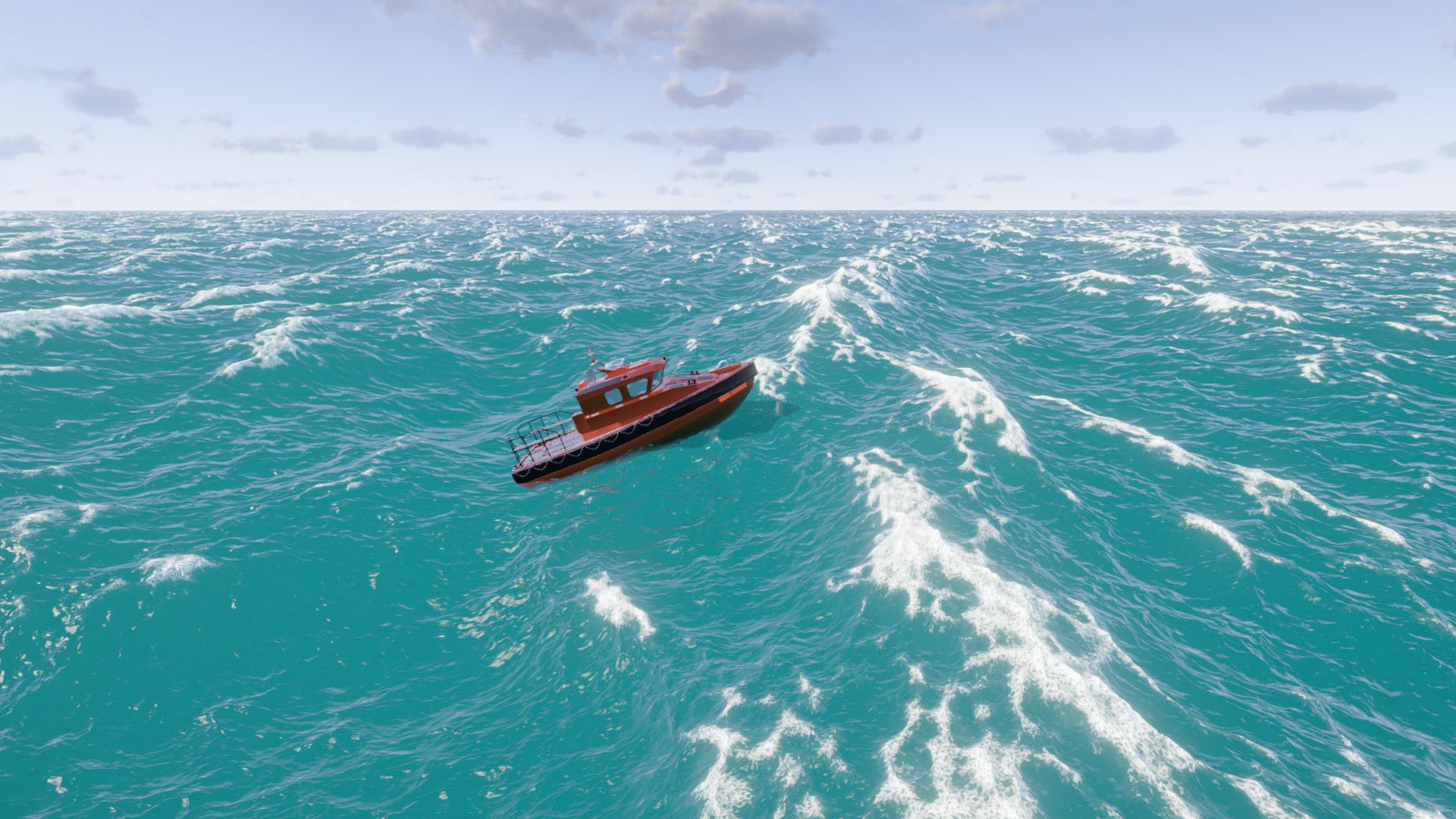}
		\label{fig:experiment1-2}
	}
	\vspace{1cm}
	\subfloat[\emph{Ten-solids} using sea state $1$.]
	{
		\includegraphics[width=0.4\textwidth]{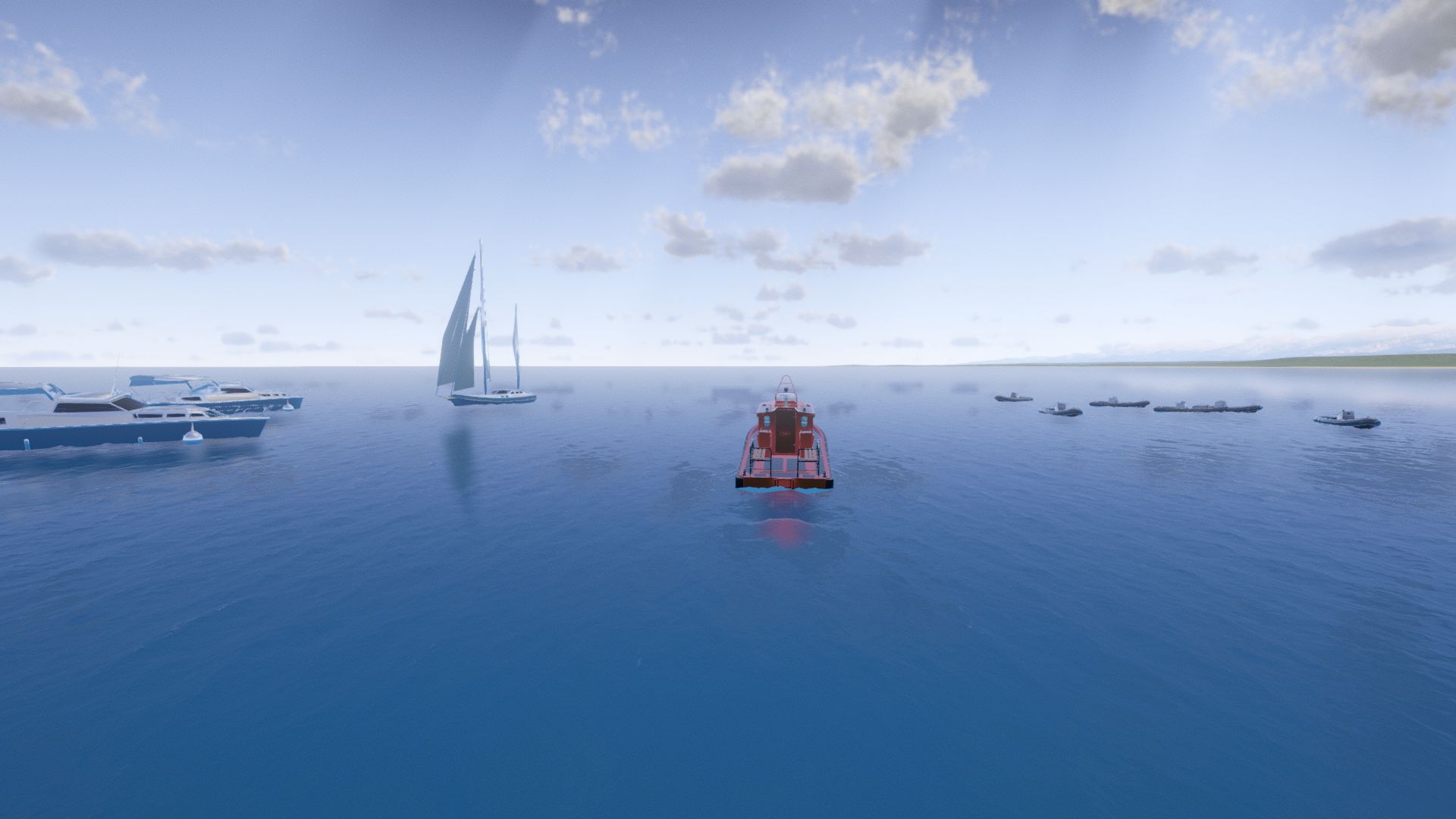}
		\label{fig:experiment2-1}
	}
	\hspace{1cm}
	\subfloat[\emph{Ten-solids} using sea state $4$.]
	{
		\includegraphics[width=0.4\textwidth]{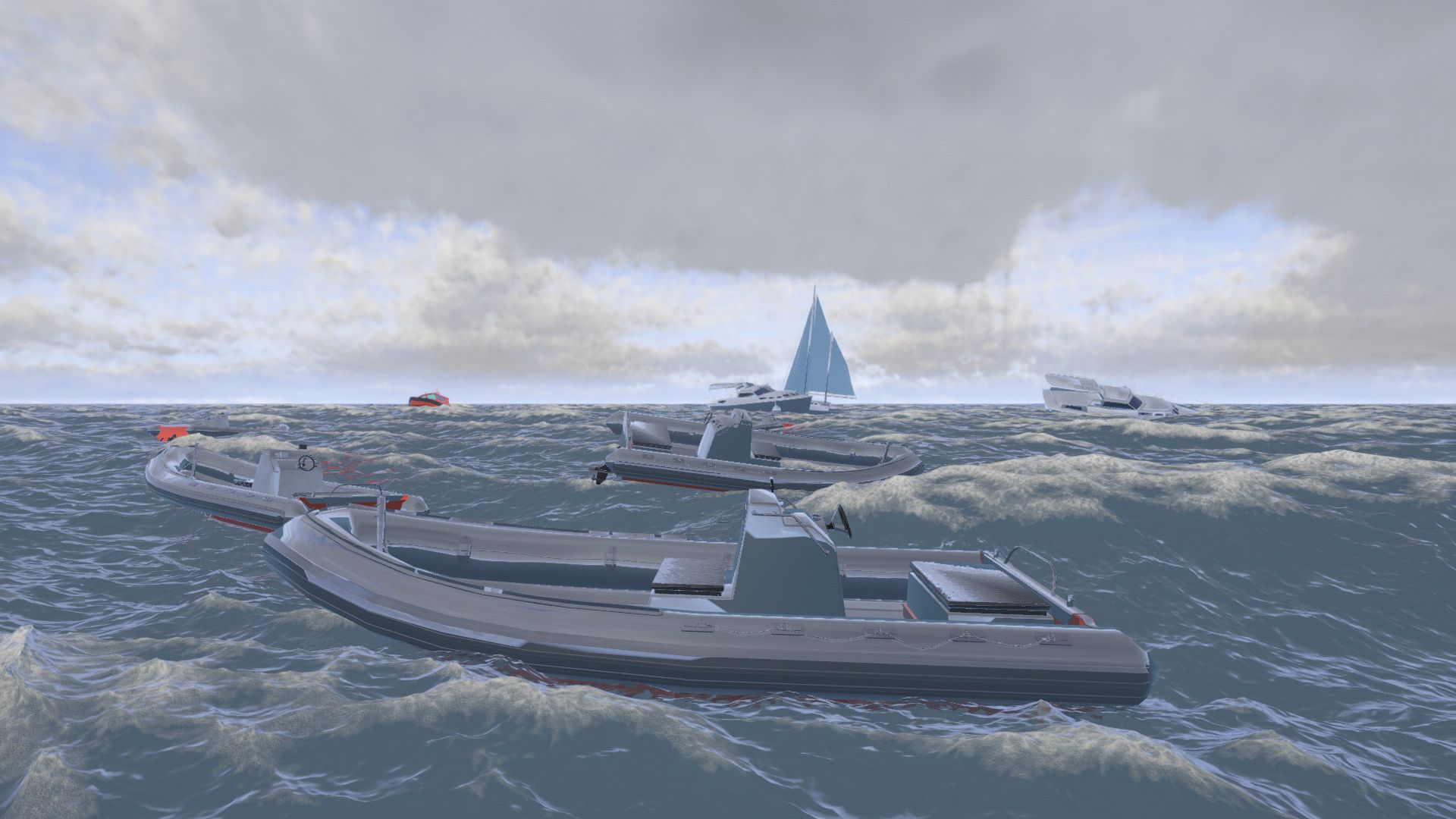}
		\label{fig:experiment2-2}
	}
	\caption{The two experiments with different sea state: \emph{one-solid} on top row, and \emph{ten-solids} on bottom one.}
	\label{fig:experiments}
\end{figure}

The first experiment called \emph{one-solid} simulates a single solid corresponding to a motorboat.
This motorboat is running in the scene.
Its model is composed of $165$ triangles for the simulation.
Two captures of \emph{one-solid} can be seen in the top row of \figurename~\ref{fig:experiments}.
The sea-state is 1 on the Beaufort scale on \figurename~\ref{fig:experiment1-1}, leading to a very calm sea.
Hence, the interaction waves appear clearly behind the motorboat.
On the contrary, the sea-state is 6 for \figurename~\ref{fig:experiment1-2}, leading to big waves and almost completely hiding interaction waves.

The second experiment called \emph{ten-solids} simulates ten solids.
It extends the first experiment as follows:
\begin{itemize}
	\item $1$ sailing boat with a mesh made of $270$ triangles.
	\item $2$ yachts with a mesh made of $146$ triangles.
	\item $6$ zodiacs with a mesh made of $568$ triangles.
\end{itemize}
This makes a total of $4135$ triangles.
Each of these ten solids uses its own grid for the solid to fluid coupling FDM.
Two captures of the second experiment can be seen in the bottom row of \figurename~\ref{fig:experiments}.
The sea state is 1 for \figurename~\ref{fig:experiment2-1} where again small interaction waves can be seen for the motorboat (the others are not propelled).
On \figurename~\ref{fig:experiment2-2} the sea state is set to 4, leading to relatively big waves.
Using 10 boats and so FDM grids does not impact too much the computation times, as seen below.

\tablename~\ref{fig:performances} shows a summary of GPU computation times for each method and with these two experiments.
As detailed in previous sections, all heavy computations in \framework are offloaded to the GPU, while the CPU is minimally involved in physical calculations. This design choice ensures that the CPU remains available for other standard game or simulation tasks, optimizing overall system performance. Therefore, CPU computation times are not considered, as they are negligible compared to GPU ones.
First, it can be observed that real-time performance is largely achieved, even with \emph{ten-solids}. One of the most computationally intensive steps is geometry computation, which presents an opportunity for further optimization. A closer inspection suggests that refining memory access patterns and minimizing warp branch divergence in the GPU kernel could enhance performance and efficiency.
Besides, as most of the computations are performed on the GPU, it can be observed that computation times do not grow linearly with the number of solids.
Indeed, \framework processes each solid and each FDM grid simulation in parallel.
\begin{figure}[!h]
	\centering
	\begin{tabular}{|p{0.28\columnwidth}|p{0.24\columnwidth}|>{\centering\arraybackslash}p{0.14\columnwidth}|>{\centering\arraybackslash}p{0.14\columnwidth}|}
		\hline
		                                  &                                              & XP 1               & XP 2                \\
		                                  &                                              & (\emph{one-solid}) & (\emph{ten-solids}) \\
		\hline
		Tessendorf's method               & Height (\S~\ref{subsec:waterHeight})         & 0.376              & 0.376               \\
		\cline{2-4}
		(\S~\ref{sec:oceanWaves})         & Velocity (\S~\ref{subsec:velWater})          & 1.175              & 1.175               \\
		\cline{2-4}
		                                  & Total                                        & 1.551              & 1.551               \\
		\hline
		Fluid-to-solid coupling           & Geometry (\S~\ref{subsec:geomParam})         & 1.125              & 4.588               \\
		\cline{2-4}
		(\S~\ref{sec:fluidToSolidAction}) & Forces (\S~\ref{subsec:forces})              & 0.406              & 3.677               \\
		\cline{2-4}
		                                  & Total                                        & 1.531              & 8.265               \\
		\hline
		Solid-to-fluid coupling           & FDM (\S~\ref{subsec:finiteDifferenceMethod}) & 0.100              & 0.831               \\
		\cline{2-4}
		(\S~\ref{sec:solidToFluidAction}) & Mask (\S~\ref{subsec:mask})                  & 0.238              & 1.656               \\
		\cline{2-4}
		                                  & Total                                        & 0.338              & 2.487               \\
		\hline
		\rowcolor{bleuC}\textbf{Total}    &                                              & \textbf{3.414}     & \textbf{12.297}     \\
		\hline
	\end{tabular}
	\caption{%
		GPU computation times in milliseconds for the main parts of \framework.
		Only GPU execution times is shown, as CPU ones are negligible.}
	\label{fig:performances}
\end{figure}

\section{Conclusions}\label{sec:conclusions}
This article presents three methods that together produce a realistic albeit approximate simulation of the ocean, and the way used to combine them together. 
These methods are simple to implement and fully parallelizable on GPU.
\Framework is a demonstration that these methods achieve real-time performance on a desktop PC (see section \ref{sec:results_and_discussions}).

This article gives an original expression for the ocean velocity associated with the Tessendorf method, as well as the key details of its implementation. This velocity is an important input for the realistic calculation of fluid-induced forces on solids.
This also improves the stability and physical coherence of the FDM for wave generation.

The two solid/fluid coupling methods are highly efficient, but they present many physical approximations.
Hence, to increase the realism of \framework, future works will include the hybridization of a smoothed particle hydrodynamics (SPH) method for solid/fluid coupling and Tessendorf methods on a larger scale. 

\subsection*{Acknowledgements}
The first author David Algis has been supported by the \href{https://www.anrt.asso.fr/fr}{Association nationale de recherche et de technologie (ANRT)}\footnote{Contract number 2021/1157} and by the \href{https://www.magelis.org/}{Pôle Image Magélis}. We would like to thank the colleague of David Algis at Studio Nyx, particularly Jeremy Bois and David Deckeur. Thanks to Naval Group for initiating this project. We would also like to thank Yannick Privat at the \href{https://irma.math.unistra.fr/}{Institut de recherche mathématique avancée (IRMA)} which help us on the velocity calculation. Finally, we would like to thanks the reviewers and editors of the Journal of Computer Graphics Techniques (JCGT) for their valuable comments and proposals to this paper. The original publication of this paper is available on their website \url{https://jcgt.org/}.


\small
\bibliographystyle{plain}
\bibliography{bibliography}


\section*{Author Contact Information}

\hspace{-2mm}
\begin{tabular}{p{0.7\textwidth}p{0.5\textwidth}}
	David Algis\newline
	Bât. H1 - SP2MI\newline
	TSA 41123\newline
	86073 Poitiers Cedex 9\newline
	FRANCE\newline
	\href{mailto:david.algis@univ-poitiers.fr}{david.algis@univ-poitiers.fr}
\end{tabular}

\appendix

\section*{Appendices}
\section{Water velocity}
\label{sec:annexe}
The water height is deduced from system of equations described in Equations (\ref{equa:systemTessendorf}), and its solution according to Tessendorf is given by Equation (\ref{equa:heightWater}) in Section \ref{sec:oceanWaves}.
This appendix uses these equations to obtain an expression of the velocity potential and the water velocity.
To simplify the derivation, only one wave vector $\mathbf{k}$ is considered in Equation (\ref{equa:heightWater}) as Equations (\ref{equa:systemTessendorf}) are all linear.

First, the velocity potential for $y=0$ is deduced from first equation in \ref{equa:systemTessendorf}, and Equations (\ref{equa:heightWater}) and \ref{equa:heightWaterBis} as follows:
\begin{align}
	\phi(\mathbf{x},0,t)
	 & = \int \frac{\partial \phi}{\partial t}(\mathbf{x},0,t)dt \nonumber                                                                                                             \\
	 & =\int-gh(\mathbf{x},t)dt \nonumber                                                                                                                                              \\
	 & = -g\exp(i\mathbf{k}.\mathbf{\mathbf{x}})\int\left(\tilde{h}_0(\mathbf{k})\exp(i\omega(k)t)+\tilde{h}_0^*(-\mathbf{k})\exp(-i\omega(k)t)\right)dt \nonumber                     \\
	 & = \frac{ig}{\omega(k)}\exp(i\mathbf{k}.\mathbf{\mathbf{x}})\left(\tilde{h}_0(\mathbf{k})\exp(i\omega(k)t)-\tilde{h}_0^*(-\mathbf{k})\exp(-i\omega(k)t)\right)\label{equa:phiy0}
\end{align}
where $\mathbf{x}=\left[\begin{array}{cc}x&z\end{array}\right]^T$.
It is assuming that potential velocity at any depth $y$ can be written as follows:
\begin{equation}
	\phi(\mathbf{x},y,t) = \tilde{\phi}(\mathbf{k}, y, t)\exp(i \mathbf{k}\cdot \mathbf{x}).
\end{equation}
Notice that this expression is convenient as it can be computed with an inverse Fast Fourier Transform.
The function $\tilde{\phi}(\mathbf{k}, y, t)$ is defined as follows:
\begin{equation}
	\tilde{\phi}(\mathbf{k}, y, t)=\frac{ig}{\omega(k)} E(\mathbf{k},y) \left(\tilde{h}_0(\mathbf{k})\exp(i\omega(k)t)-\tilde{h}_0^*(-\mathbf{k})\exp(-i\omega(k)t)\right)
\end{equation}
where \textit{attenuation function} $E(\mathbf{k},y)$ has to be determined from Equations \ref{equa:systemTessendorf}.
A first condition comes from Equation \ref{equa:phiy0}:
\begin{equation}
	E(\mathbf{k}, 0)=1 \label{equa:initialE}
\end{equation}
Then, the second equation from system \ref{equa:systemTessendorf} can be imposed on the attenuation function.
First, the left term is developed as follows:
\begin{align}
	\Delta \phi\nonumber= & \left(-k_x^2 E(\mathbf{k},y) -k_z^2E(\mathbf{k},y) + E''(\mathbf{k},y)\right)\nonumber                                                                                      \\
	                      & \times \frac{ig}{\omega(k)}\left(\tilde{h}_0(\mathbf{k})\exp(i\omega(k)t)-\tilde{h}_0^*(-\mathbf{k})\exp(-i\omega(k)t)\right)
	\exp\left( i \mathbf{k}\cdot \mathbf{x}\right)\nonumber                                                                                                                                             \\
	=                     & \left(E''(\mathbf{k},y)-k^2E(\mathbf{k},y)\right) \nonumber                                                                                                                 \\
	                      & \times \frac{ig}{\omega(k)}\left(\tilde{h}_0(\mathbf{k})\exp(i\omega(k)t)-\tilde{h}_0^*(-\mathbf{k})\exp(-i\omega(k)t)\right) \exp\left(i \mathbf{k}\cdot \mathbf{x}\right)
\end{align}
Hence, condition $\Delta \phi=0$ is equivalent to find a solution of the following second order ordinary differential linear equation:
\begin{equation}
	E''(\mathbf{k}, y)-k^2E(\mathbf{k}, y) = 0.
\end{equation}
It can be deduced that for $-H<y\leq 0$, $E(\mathbf{k}, y)$ is as follows:
\begin{equation}
	E(\mathbf{k}, y)=C_1\cosh(ky+C_2)
\end{equation}
where $C_1,C_2\in\mathbb{R}$ are constant.
The derivative regarding height $y$ of $\phi(\mathbf{x},-H,t)$ can be rewritten as follows:
\begin{align*}
	\frac{\partial\phi(\mathbf{x},y,t)}{\partial y}= &
	\frac{\partial E(\mathbf{k},y)}{\partial y}\frac{ig}{\omega(k)}\left(\tilde{h}_0(\mathbf{k})\exp(i\omega(k)t)-\tilde{h}_0^*(-\mathbf{k})\exp(-i\omega(k)t)\right)\exp\left(i\mathbf{k}\cdot\mathbf x\right)               \\
	=                                                & kC_1\sinh(ky+C_2)                                                                                                                                                      \\
	                                                 & \times\frac{ig}{\omega(k)}\left(\tilde{h}_0(\mathbf{k})\exp(i\omega(k)t)-\tilde{h}_0^*(-\mathbf{k})\exp(-i\omega(k)t)\right)\exp\left(i\mathbf{k}\cdot\mathbf x\right)
\end{align*}
Hence, at depth $y=-H$ the fourth equation of system \ref{equa:systemTessendorf} leads to:
\begin{align}
	\sinh(-kH+C_2) & = 0 \nonumber \\
	C_2            & = kH.
\end{align}
The constant $C_1$ is obtained using Equation~\ref{equa:initialE} as follows:
\begin{align}
	E(\mathbf{k},0) & =1\nonumber           \\
	C_1\cosh(C_2)   & =1\nonumber           \\
	C_1             & =\frac{1}{\cosh(kH)}.
\end{align}
To resume, the full expression of the attenuation function for $-H\leq y\leq 0$ is as follows:
\begin{equation}
	E(\mathbf{k}, y)=\frac{\cosh\left(ky+kH\right)}{\cosh(kH)}.
\end{equation}
However, this expression cannot be used for large values of $k$ and sea height $H$, where the hyperbolic cosine becomes bigger than the limit of real numbers even in double precision.
Hence, $E$ quickly reach the indeterminate form $\frac{\infty}{\infty}$.
To fix this problem, the following physical assumption is made: it is assumed \textbf{deep water}, \ie that the sea depth $H$ tends towards $\infty$.
First, $E$ is rewritten as follows:
\begin{align*}
	E(\mathbf{k},y) & = \frac{\cosh(ky+kH)}{\cosh(kH)}\nonumber                            \\
	                & = \frac{\cosh(ky)\cosh(kH) + \sinh(ky)\sinh(kH)}{\cosh(kH)}\nonumber \\
	                & = \cosh(ky) + \sinh(ky)\tanh(kH)
\end{align*}
Then, considering that $\underset{H \rightarrow \infty}{\lim}\tanh(kH)=1$, the following approximation is obtained:
\begin{align}\label{equa:attenuationDeep1}
	E(\mathbf{k},y) \approx & \underset{H \rightarrow \infty}{\lim} E(\mathbf{k},y)\nonumber               \\
	\approx                 & \underset{H \rightarrow \infty}{\lim}\cosh(ky) + \sinh(ky)\tanh(kH)\nonumber \\
	\approx                 & \exp(ky).
\end{align}

The third equation of the system \ref{equa:systemTessendorf} gives the following equality:
\begin{equation}
	\label{equa:derivHeqDerivPhi}
	\frac{\partial h}{\partial t}(\mathbf{k},t)=\frac{\partial\phi}{\partial y}(\mathbf{k}, 0, t)
\end{equation}
Once again, we only considered one wave vector $\mathbf{k}$ as Equations are all linear, in consequence Equation \ref{equa:derivHeqDerivPhi} becomes:

\begin{align}
	\frac{\partial \tilde{h}}{\partial t}(\mathbf{k},t) - \frac{\tilde{\partial}\phi}{\partial y}(\mathbf{k}, 0, t) & =0 \nonumber                                \\
	i\omega(k)\left(\tilde{h}_0(\mathbf{k})\exp(i\omega(k)t)-\tilde{h}_0^*(-\mathbf{k})\exp(-i\omega(k)t)\right) - \frac{igk}{\omega(k)} E(\mathbf{k},0)\nonumber \\
	\times \left(\tilde{h}_0(\mathbf{k})\exp(i\omega(k)t)-\tilde{h}_0^*(-\mathbf{k})\exp(-i\omega(k)t)\right)       & = 0 \nonumber                               \\
	\omega(k)- \frac{gk}{\omega(k)}                                                                                 & = 0 \nonumber                               \\
	\omega(k) = \sqrt{gk} \label{equa:relationDispersionDeduce}
\end{align}

This last Equation \ref{equa:relationDispersionDeduce} gives the expression of the dispersion relation used in the Section \ref{sec:oceanWaves}.

Equation~\ref{equa:attenuationDeep1} and the definition of potential velocity $\nabla\phi=\mathbf{v}$ lead to the velocity $\mathbf{v}$. First, function $\tilde{\phi}$ is rewritten as follows:
\begin{equation}
	\tilde{\phi}(\mathbf{k}, y, t)=\frac{ig}{\omega(k)} E(\mathbf{k},y) \left(\tilde{h}_0(\mathbf{k})\exp(i\omega(k)t)-\tilde{h}_0^*(-\mathbf{k})\exp(-i\omega(k)t)\right).
\end{equation}
Then the gradient of $\phi$ is as follows:
\begin{equation}
	\nabla \phi = \begin{pmatrix}
		v_x \\
		v_y \\
		v_z
	\end{pmatrix} =\tilde{\mathbf{v}}(\mathbf{k}, y, t)\exp\left(i\mathbf{k}\cdot\mathbf{x}\right)
\end{equation}
where $\tilde{\mathbf{v}}(\mathbf{k}, y, t)$ is the vector equal to:
\begin{equation}
	\tilde{\mathbf{v}}\left(\mathbf{k}, y, t\right)=%
	E\left(\mathbf{k},y\right)\left(\tilde{h}_0\left(\mathbf{k}\right)\exp\left({i\omega(k)t}\right)-\tilde{h}_0^*\left(-\mathbf{k}\right)\exp\left({-i\omega(k)t}\right)\right)
	\left[
		\begin{array}{c}
			\displaystyle\frac{-k_xg}{\omega(k)} \\
			\displaystyle i\omega(k)             \\
			\displaystyle\frac{-k_zg}{\omega(k)}
		\end{array}
		\right],
\end{equation}

Finally, reintroducing a sum on a set of waves vectors $\mathbf{k}$ leads to:
\begin{equation}
	\nabla \phi =
	\sum_{\mathbf{k}}\tilde{\mathbf{v}}(\mathbf{k}, y, t)\exp\left(i\mathbf{k}\cdot\mathbf{x}\right).
\end{equation}

Note that the expression of $E\left(\mathbf{k},y\right)$ (see Equation \ref{equa:attenuationDeep1}) is valid for height $y\leq0$, but not usable above for $y>0$.
Several possibilities exist to overcome this problem:
\begin{enumerate}
	\item The most naive one assumes $E(\mathbf k,y>0)=1$.
	\item A more realistic solution is based on a linear extrapolation using the derivative at $y=0$, as follows:
	      \begin{equation}
		      \label{equa:attenuationExtrapolation}
		      E(\mathbf{k}, y)=
		      \begin{cases}
			      E(\mathbf k,0)+y\frac{\partial E}{\partial y}(\mathbf k,0) & \text{if~}y>0, \\
			      \exp(ky)                                                   & \text{else.}
		      \end{cases}
	      \end{equation}
\end{enumerate}

More sophisticated solutions are summarized in the Molin's book (2002) \cite{molinHydrodynamiqueStructuresOffshore2002}.
Nevertheless, they are too complex to compute as they make the attenuation function depending on $h(\mathbf x,t)$.
For this reason, \framework uses Equation~\ref{equa:attenuationExtrapolation}, with the following simplified form:
\begin{equation}
	\label{equa:attenuationFull}
	E(\mathbf{k}, y)=
	\begin{cases}
		1+ky     & \text{if~}y>0, \\
		\exp(ky) & \text{else.}
	\end{cases}
\end{equation}

\section{Navier-Stokes equations approximation}\label{sec:annexe-NS}
This appendix proposes a derivation of Tessendorf's equations \ref{equa:systemTessendorf} from Navier-Stokes equations. As stated in Teman (2001) \cite{temamNavierStokesEquations2001} and assuming conservation of mass, Navier-Stokes equations are given by:
\begin{equation}
	\begin{cases}
		\frac{\partial}{\partial t}\left(\rho \mathbf{u}\right) + \mathbf{u} \cdot \nabla (\rho \mathbf{u}) = - \nabla p + \rho\mathbf{g} +  \nabla^2 (\rho\nu \mathbf{u}) \\
		\frac{\partial}{\partial t}\left(\rho \mathbf{u}\right) + \nabla\cdot (\rho\mathbf{u}) = 0
	\end{cases}
	\label{equa:ns0}
\end{equation}
where $\mathbf{u} :\mathbb{R}^3\mapsto \mathbb{R}^3$ is the flow velocity, $\rho$ is the time-dependent density of the fluid, $p: \mathbb{R}^3\mapsto \mathbb{R}$ is the pressure of the fluid, $\nu$ is the dynamic viscosity of the fluid.
The first equation is called \emph{Cauchy momentum equation} and the second \emph{continuity equation}.

Boundary conditions are necessary for simulations. In the case of solid, the boundary equation is as follows:
\begin{equation}
	\mathbf{u}\cdot\mathbf{n} = \mathbf{u}_{solid} \cdot \mathbf{n}
\end{equation}
where, $\mathbf{u}_{solid}$ is the velocity of the solid and $\mathbf{n}$ is the normal to its surface.

Some assumptions are made to simplify these equations, and to derive Tessendorf equations.
They are discussed below.

\paragraph{Incompressibility}
As mention in Section~3.3.2 of Hulin \etal (2001) \cite{hulinHydrodynamiquePhysique2001}, a fluid can be considered incompressible if $u<<c_{sound}$ where $c_{sound}$ if the sound speed in ocean which is approximately equal\footnote{See \url{https://en.wikipedia.org/wiki/Sound_speed_profile} to $c_{sound} = 1530~m.s^{-1}$.}.
Aside from phenomena such as hurricane, this inequality seems reasonable for oceans, and leads to consider that ocean is incompressible.
More formally:
\begin{equation}
	\rho(t)=\rho.
\end{equation}
Consequently, density can be factorized in each term of equations \ref{equa:ns0}.

\paragraph{Inviscid}
A fluid can be considered inviscid if $Re>>1$ (see Section~2.3.1 of Hulin \etal (2001) \cite{hulinHydrodynamiquePhysique2001}) where $Re=VL/\nu$ is the \emph{Reynolds number} of the fluid defined by the order of magnitude of its velocity $V$, its characteristic length $L$ and its dynamic viscosity $\nu$.
For Tessendorf derivation, $V$ is the velocity of waves, $L$ its wave length and for seawater at $20$°C its is stated that $\nu=1.0508.10^{-6}$ (see the measure of the ITTC \cite{ittcFreshWaterSeawater2011}).
Therefore, the inequality is true for product $VL >> 10^{-5}$.
For these values the ocean can be considered as flat.
Hence, the viscosity term $\nu$ of Navier-Stokes equations \ref{equa:ns0} can be neglected.

\paragraph{Irrotational}
If a fluid is irrotational at time $t_0$, it must remain irrotational unless there are changes in the boundary conditions.
Clearly, flat sea are irrotational fluid, corresponding to $\nabla\times\mathbf{u} = \mathbf{0}$.
Therefore, it can assume that for calm waves the fluid remains irrotational.
Then, there exists a scalar field called \textbf{velocity potential} defined by $\phi = \mathbf{\nabla} \phi$ from \emph{Helmholtz decomposition}\footnote{See \url{https://en.wikipedia.org/wiki/Helmholtz_decomposition}.}.

These three assumptions lead to the Bernoulli equation (see Section~5.3.2 of Hulin \etal (2001), which is the following simplified version of the Cauchy momentum equation:
\begin{equation}
	\frac{\partial \phi}{\partial t} + \frac{1}{2}\|\nabla\phi\|^2 + \frac{p}{\rho} + gy = 0
	\label{equa:bernoulli}
\end{equation}

Moreover, the continuity equation is now:
\begin{equation}
	\Delta \phi = 0 \label{equa:continuityS}
\end{equation}

Nevertheless, these non-linear equations are still too complex to be solved at large scaled and in real-time.

\paragraph{Interface condition}
It is necessary to define a boundary condition at the interface between the ocean fluid and the air fluid.
To simplify, it is assumed that the air can be represented as a region with constant pressure\footnote{As shown by the \href{https://en.wikipedia.org/wiki/Barometric_formula}{barometric formula} the atmospheric pressure has a very low evolution in function of altitude.}.
Since air density is a hundred of times lighter than water, its effects on the free surface are assumed to be negligible.
As with an incompressible fluid, only the differences in pressure matter.
Therefore, the air pressure is defined as an arbitrary constant, leading to:
\begin{equation}
	p = 0\text{~~~~~~~~~~~at free surface.}
\end{equation}

Hence, this allows to remove the pressure term in Bernoulli equation \ref{equa:bernoulli} at free surface.

\paragraph{Height field}
At this point, equations are still too complex to be used in real-time, especially in terms of dimension:
it seems inconceivable to solve them at large scale in 3D.
Therefore, to reduce their dimensions it is assumed that the surface geometry is described by a height field, as follows:
\begin{equation}
	y = h(\mathbf x, t).
\end{equation}

This hypothesis excludes a lot of physical phenomena like breaking waves, but it helps to reach large scale domain.

\paragraph{Free surface follows the velocity of water}
Next step assumes that the free surface follows the velocity of the water.
Hence, for all $\mathbf x$ on the horizontal axes at time $t + dt$ the height field is as follows:
\begin{equation}
	h(\mathbf{x^{\prime}}, t + dt) = h(\mathbf x, t) + {v_y}dt\label{equa:freeSurfaceFollowWater0}
\end{equation}
where:
\begin{equation}
	\mathbf{x^{\prime}} = \mathbf x + dt\left[\begin{array}{c}v_x\\ v_z\end{array}\right]
	\label{equa:freeSurfaceFollowWater1}
\end{equation}

The Taylor expansion of $h(\mathbf{x^{\prime}}, t + dt)$ leads to:
\begin{equation}
	h(\mathbf{x^{\prime}}, t + dt) = h(\mathbf{x}, t + dt) + v_xd_t\frac{\partial h(\mathbf{x}, t)}{\partial x} + v_zd_t\frac{\partial h(\mathbf{x}, t)}{\partial z} + \ldots
\end{equation}

Neglecting the nonlinear terms in the Taylor expansion and reintegrating the development of $h(\mathbf{x^{\prime}}, t + dt)$ into Equation \ref{equa:freeSurfaceFollowWater0} and then dividing by $dt$ leads to the following result:
\begin{equation}
	\begin{aligned}
		\frac{h(\mathbf x, t + dt) - h(\mathbf x, t)}{dt} + v_x\frac{\partial h(\mathbf x, t)}{\partial x} + v_z\frac{\partial h(\mathbf x, t)}{\partial z} & = v_y \\
		\frac{\partial h}{\partial t} + v_x\frac{\partial h}{\partial x} + v_z\frac{\partial h}{\partial z}                                                 & = v_y
	\end{aligned}
\end{equation}

This leads to the following \emph{advection equation}:
\begin{equation}
	\frac{D}{Dt}(h(\mathbf x) - y) = 0
	\label{equa:advection}
\end{equation}

\paragraph{Remove quadratic term}
The quadratic term in the Bernoulli equation \ref{equa:bernoulli} makes solving the equation too complex.
Therefore, the velocity is assumed to be sufficiently small to neglect the quadratic term relative to the others.
It leads to the “final” Bernoulli equation expressed as follows:
\begin{equation}
	\frac{\partial \phi}{\partial t} = -gh(\mathbf x,t) \label{equa:bernoulliS}
\end{equation}

\paragraph{Smooth free surface}
Similarly, the nonlinear terms in the advection equation \ref{equa:advection} make the equation's resolution too complex.
So the free surface height is assumed to be sufficiently smooth to neglect these terms relative to others.
As a result, advection equation \ref{equa:advection} becomes:
\begin{equation}
	\frac{\partial h}{\partial t} = \frac{\partial \phi}{\partial y} \label{equa:advectionS}
\end{equation}

\paragraph{Flat ocean bottom}
Now, it is assumed that the ocean depth is constant, and equal to $H$.
Then, the ocean floor is stationary and its normal to the ocean floor is $(0,1,0)$.
Hence, boundary conditions at $y=-H$ are defined as:
\begin{equation}
	\begin{aligned}
		\mathbf{u}\cdot\mathbf{n}        & = 0 \\
		\frac{\partial \phi}{\partial y} & = 0
	\end{aligned}
	\label{equa:bottomCondition}
\end{equation}

\paragraph{Infinite ocean}
To avoid boundary conditions on $\mathbf x$, it is assumed that the ocean is horizontally infinite and periodic, allowing a finite simulation area.
Therefore, the simulation area size is defined squared as $L \times L$.
It leads to an ocean tile that can be infinitely duplicated.

\paragraph{Small free surface}
To simplify the location at which all equations are solved, it is assumed that the height of the free surface is sufficiently low. Hence, the equations can be solved at $\mathbf x=\mathbf0$.

\paragraph{Conclusion}
Finally, putting all together leads to the Tessendorf solution.
More precisely, Tessendorf equations system is made from Equations (\ref{equa:bernoulliS}), (\ref{equa:continuityS}), (\ref{equa:advectionS}), (\ref{equa:bottomCondition}).

\section{Two Fourier transforms in one}
\label{sec:annexe-2IFFT}
This section proposes a proof of Theorem \ref{theorem:ifft}.

\begin{proof}
	This demonstration relies on the following formula that links the real part of the IFFT and the sum of the coefficients:
	\begin{equation}
		\label{equa:realFormula}
		\mathcal{F}^{-1}\left(\frac{1}{2}\left(A+A^{\dagger}\right)\right) =
		\Re\left(\mathcal{F}^{-1}\left(A\right)\right)
	\end{equation}
	where $A=(a_{n,m})$ is any Hermitian complex matrix of $M_n(\mathbb{C})$, and $A^{\dagger}$ the transpose of its conjugate.
	Using the definition and linearity of the discrete Fourier transform, it follows for one term $(k,l)$ of the IFFT:
	\begin{align}
		\mathcal{F}^{-1}\left(A+A^{\dagger}\right)_{k,l} & =
		\mathcal{F}^{-1}\left(A\right)_{k,l} + \mathcal{F}^{-1}\left(A^{\dagger}\right)_{k,l}\nonumber                                                                                                                  \\
		                                                 & =\sum_{n,m}a_{n,m}\exp\left(-i\frac{2\pi}{N}\left(kn+lm\right)\right) + \sum_{m,n}a_{m,n}^{\ast}\exp\left(-i\frac{2\pi}{N}\left(km+ln\right)\right)\nonumber \\
		                                                 & =\sum_{n,m}(a_{n,m}+a_{n,m}^{\ast})\exp\left(-i\frac{2\pi}{N}\left(kn+lm\right)\right)\nonumber                                                              \\
		                                                 & = 2\Re\left(\mathcal{F}^{-1}\left(A\right)_{k,l}\right)
	\end{align}

	Since $A$ is Hermitian by hypothesis, Equation (\ref{equa:realFormula}) becomes:
	\begin{equation}
		\mathcal{F}^{-1}\left(A\right) = \Re\left(\mathcal{F}^{-1}\left(A\right)\right)
	\end{equation}

	Applying this last equation to $X$ and $iY$ leads to:
	\begin{equation}
		\begin{aligned}
			\mathcal{F}^{-1}\left(X+iY\right) & =
			\mathcal{F}^{-1}\left(X\right) + i\mathcal{F}^{-1}\left(Y\right)                                                                       \\
			                                  & = \Re\left(\mathcal{F}^{-1}\left(X\right)\right) + i\Re\left(\mathcal{F}^{-1}\left(Y\right)\right)
		\end{aligned}
	\end{equation}

\end{proof}

\section{Evaluation of the Number of Iterations for Convergence}
\label{sec:evaluation_of_the_number_of_iterations_for_convergence}

The optimal number of iterations ($N_{\text{iter}}$) required for the iterative retrieval of water height was determined through two experiments, focusing on accuracy and computational efficiency.

\begin{figure}[h!]
	\centering
	\includegraphics[width=0.9\textwidth]{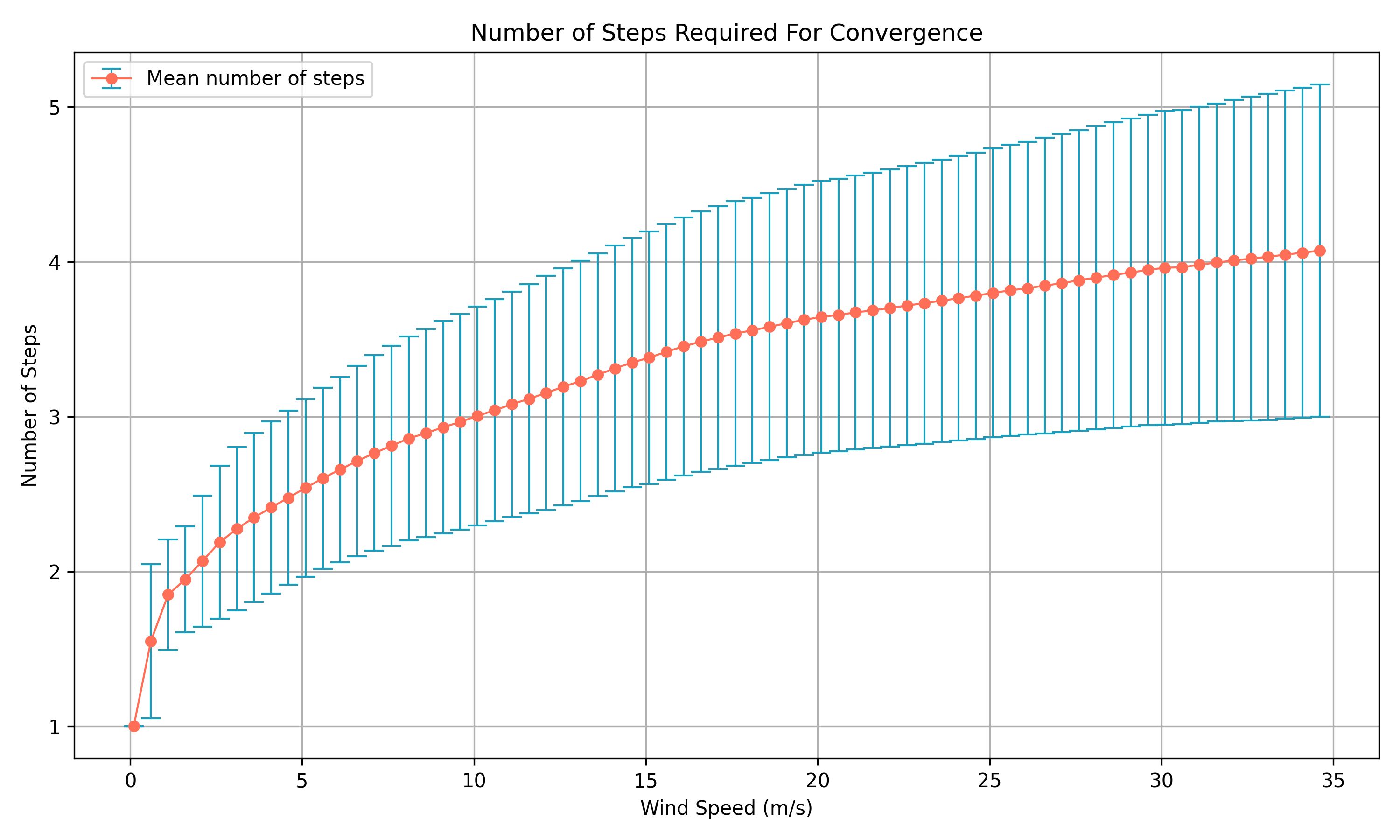}
	\caption{Mean number of steps required for convergence as a function of wind speed. Error bars represent the variance of iterations across $1000$ test points.}
	\label{fig:convergence_steps}
\end{figure}

The first experiment aimed to evaluate the number of iterations required for the water height computation to converge within a given precision. This test has been conducted on different wind speeds as it makes the height of water significantly. Wind speeds ranged from $0.1$ m/s to $35$ m/s in $0.5$ m/s increments, with convergence tested at $10^3$ random positions uniformly distributed over a $10^4\times 10^4$ m domain. The iterative process was repeated until the difference between consecutive height estimates was smaller than $0.01$ m. The results, shown in Figure~\ref{fig:convergence_steps}, indicate that the required number of iterations increases with wind speed but stabilizes at approximately $N_{\text{iter}} = 4$ for higher wind speeds.

\begin{figure}[h!]
	\centering
	\includegraphics[width=0.9\textwidth]{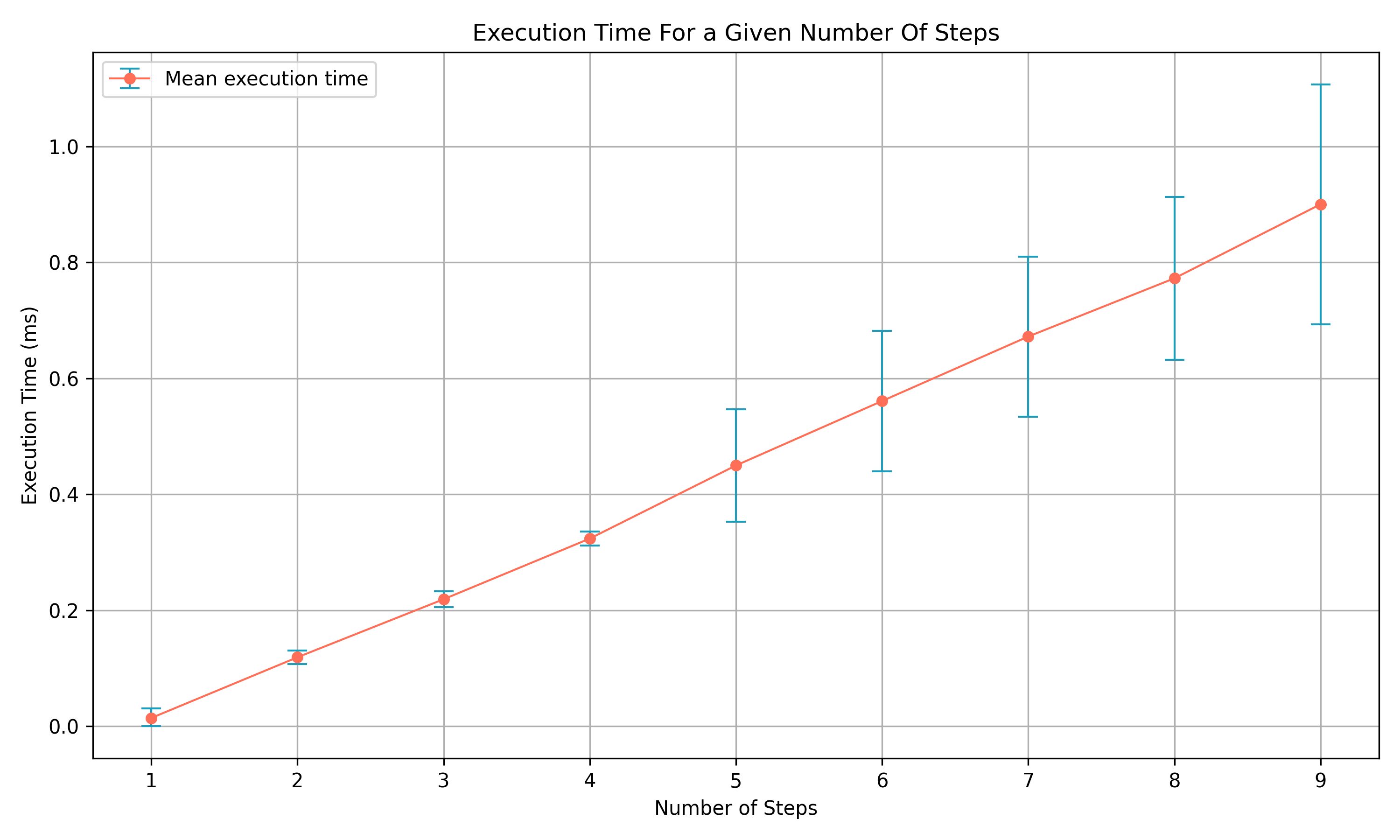}
	\caption{Average execution time per iteration count over $1000$ tests, measured across $10^5$ sample positions. Error bars indicate variance in execution time.}
	\label{fig:execution_time}
\end{figure}

The second experiment aimed to measure the computational cost associated with varying numbers of iterations. The execution time was assessed using a CUDA-based implementation, where the iterative retrieval of water height was performed for different values of $N_{\text{iter}}$. Each test was executed $10^3$ times on a GPU, and execution times were recorded across $10^5$ randomly sampled positions. The measurements included the minimum, maximum, mean, and variance of execution times for each iteration count. The results, presented in Figure~\ref{fig:execution_time}, show that while execution time increases with the number of iterations, the rate of increase diminishes after four iterations. This confirms that $N_{\text{iter}} = 4$ provides an optimal balance between computational efficiency and accuracy, ensuring reliable convergence across tested conditions.

\end{document}